%% file: main.tex
\documentclass[12pt]{article}
\usepackage[utf8]{inputenc}
\usepackage{float}
\usepackage{authblk}
\usepackage{amsmath}
\usepackage{graphicx}
\usepackage{paralist}
\usepackage{mathtools}
\usepackage{fullpage}
\usepackage{xcolor}
\usepackage[font=small]{caption} %
\usepackage{subcaption}
\usepackage{booktabs}
\usepackage{ccis-macros}
\usepackage{makecell}
\usepackage{enumitem}
\usepackage{changepage}
\usepackage{placeins}
\usepackage{hyperref}

\usepackage[
backend=biber,
style=apa,
sorting=none,
citestyle=numeric-comp,
natbib=true]{biblatex}

\AtEveryBibitem{%
  \clearfield{issn} %
  \clearfield{isbn} %
  \clearfield{annotation} %
}

\DeclareSourcemap{
  \maps[datatype=bibtex]{
    \map{
      \pertype{misc}
      \step[fieldsource=howpublished, final]
      \step[fieldset=url, origfieldval, final]
      \step[fieldset=howpublished, null]
    }
  }
}

\DeclareFieldFormat{labelnumberwidth}{\mkbibbrackets{#1}}
\defbibenvironment{bibliography}
  {\list
     {\printtext[labelnumberwidth]{%
      \printfield{labelprefix}%
      \printfield{labelnumber}}}
     {\setlength{\labelwidth}{\labelnumberwidth}%
      \setlength{\leftmargin}{\labelwidth}%
      \setlength{\labelsep}{\biblabelsep}%
      \addtolength{\leftmargin}{\labelsep}%
      \setlength{\itemsep}{\bibitemsep}%
      \setlength{\parsep}{\bibparsep}}%
      }
  {\endlist}
  {\item}

{\small \bibliography{refs}}

\hypersetup{
    colorlinks,
    linkcolor={red!50!black},
    citecolor={blue!50!black},
    urlcolor={blue!80!black}
}

\makeatletter 
\renewcommand\AB@affilsepx{, \protect\Affilfont}
\makeatother

\title{Engagement Outweighs Exposure to Partisan and Unreliable News within Google Search}

\author[1,2]{Ronald E. Robertson}
\author[2]{Jon Green}
\author[2]{Damian J. Ruck}
\author[3]{\\Katherine Ognyanova} 
\author[2]{Christo Wilson}
\author[2]{David Lazer}

\affil[1]{Stanford University}
\affil[2]{Northeastern University}
\affil[3]{Rutgers University}

\date{}

\usepackage{titling}
\begin{document}

\phantomsection
\addcontentsline{toc}{subsection}{Abstract}

\maketitle
\input{abstract}

\newpage

\phantomsection
\addcontentsline{toc}{subsection}{Introduction}

The prevalence of partisan and unreliable online news is a topic of ongoing concern among policymakers, civil society organizations, and academics~\citep{lazer2018science,watts2021measuring,pennycook2021shifting,lewandowsky2017misinformation}.
These concerns often center the role of online platforms, such as search engines or social media, in directing people to such content, and the societal impacts that such guiding may incur~\citep{wagner2021measuring,rahwan2019machine}.
The theoretical grounding for such concerns generally involves \textit{selective exposure}---the tendency to choose political information that is congruent with one's existing beliefs~\citep{taber2006motivated, iyengar2008selective, iyengar2009red}---and its counterparts: \textit{echo chambers}~\citep{sunstein2001republic,cinelli2021echo} and \textit{filter bubbles}~\citep{pariser2011filter, bakshy2015exposure}.

In online settings, echo chambers often center the role of users' choices, including direct navigation, search query formulation, or social networking decisions,
and how they can create settings ``in which most available information conforms to pre-existing attitudes and biases''~\cite{lewandowsky2017misinformation}.
In contrast, filter bubbles center the role of algorithmic curation, such as the production of a social media feed or a search results page,
where content ``selected by algorithms according to a viewer's previous behaviors'' can create a feedback loop that limits exposure to cross-cutting content~\citep{bakshy2015exposure}.
Although definitions for these concepts vary and overlap, both can be described in terms of user choice and algorithmic curation~\cite{cardenal2019digital,fletcher2021more}, and both predict a similar observable outcome: partisans will see and select a significant amount of identity-congruent content.

Recent research on partisan and unreliable online news has primarily focused on users' choices on social media platforms~\citep{pennycook2021shifting, cinelli2021echo,johnson2020online,hosseinmardi2021examining,chen2021neutral} or during general web browsing~\citep{guess2020exposure, cardenal2019digital, allen2020evaluating,muise2022quantifying,garimella2021political}, leaving open questions about algorithmic curation, especially on web search engines.
The importance of studying news in web search is evident from long-standing concerns about the impact of search engines~\citep{introna2000shaping,lawrence1999accessibility,metaxas2005weba,vaidhyanathan2011googlization}, and urgent in light of several recent findings.
For example, survey and digital trace studies have found that web search plays a central role in directing users to online news~\citep{allcott2017social, guess2020exposure, bentley2019understanding, newman2019digital, mitchell2017how,edelman2021edelman}, qualitative work has documented patterns of unreliable and false information in web search results~\cite{golebiewski2019data,noble2018algorithms}, and lab experiments suggest politically-biased search rankings can influence political opinions~\cite{epstein2015search,epstein2017suppressing}.
However, research on news seen within web search has generally been limited to algorithm auditing studies---where what real users might have seen in their search results was estimated using simulated user behavior (hypothetical exposure)~\citep{hannak2013measuring, metaxa2019search, robertson2018auditinga, trielli2020partisan, fischer2020auditing,kawakami2020media}---and digital trace studies---where what real users might have seen was estimated from available click logs (selected exposure)~\citep{guess2020exposure,wojcieszak2021avenues,fletcher2021more,cardenal2019digital,peterson2021partisan}.

We operationalize the two sides of a user-platform interaction as exposure and engagement, defining \textit{exposure} as the URLs people see while visiting a platform and \textit{engagement} as the URLs that people interact with while on that platform or while browsing the web more broadly. 
We also define \textit{follows} as the intersection of these data types---engagement conditional on exposure---representing instances in which a person engages with a URL immediately after visiting a platform.
These distinctions build on similar terms used in previous research by incorporating a wider variety of on- and off-platform web behaviors (see Methods).
For example, \citet{bakshy2015exposure} used internal Facebook data to measure exposure as cases ``in which a link to the content appears on screen in an individual's NewsFeed,'' and described on-platform engagement as ``clicks'' and ``consumption,'' but did not measure off-platform behavior.
To date, studies involving ecologically valid exposure and engagement data---what real users saw and did during their everyday use of a platform---have largely been limited to internal studies conducted by social media platforms~\cite{bakshy2015exposure,huszar2022algorithmic}.

Here we advance research on user choice and algorithmic curation through a two-wave study in which we deployed a custom web browser extension to collect ecologically valid measures of both exposure and engagement on Google Search.
During the 2018 and 2020 US election cycles, we invited participants to complete a survey and voluntarily install our extension with informed consent.
We then merged those exposure and engagement data with domain-level (\eg \texttt{bbc.com}) measures of partisan and unreliable news and used an unsupervised method on the text of participants' queries to quantify differences in their search behavior.
Paired with the surveys, these data enabled us to examine differences among groups with characteristics, such as partisan identification and age, that have previously been linked to greater interaction with partisan or unreliable news~\cite{grinberg2019fake,guess2020exposure,guess2019less}.

Results from both study waves show that participants' partisan identification had a small and inconsistent relationship with the amount of partisan and unreliable news they were exposed to on Google Search, a more consistent relationship with the search results they chose to follow, and the most consistent relationship with their overall engagement.
Differences in participants' demographic characteristics and the content of their search queries largely explained the small differences we found for exposure to partisan and unreliable news on Google Search, suggesting an absence of filter bubbles in this context.
However, the more consistent differences we observed in participants' follows and overall engagement with partisan and unreliable news, representing their on- and off-platform choices, suggest at least some degree of online echo chambers.
These findings shed light on the role of Google Search in leading its users to partisan and unreliable news, highlight the importance of measuring both user choice and algorithmic curation when studying online platforms, and are consistent with prior work on general web browsing~\cite{cardenal2019digital,fletcher2021more} and Facebook's NewsFeed~\cite{bakshy2015exposure}.
Last, and also consistent with studies on social media and general web browsing~\cite{guess2020exposure,guess2019less,grinberg2019fake}, we found that engagement with unreliable news is generally rare and concentrated among a small number of older participants who identify as strong Republicans.

\subsection{Collecting Exposure and Engagement Data}

From October through December 2018, and again in April through December 2020, we recruited participants to take a survey and optionally install a custom browser extension we made for Chrome and Firefox.
In the survey, participants self-reported both their seven-point partisan identification (7-point PID; strong Democrat, not very strong Democrat, lean Democrat, Independent, lean Republican, not very strong Republican, strong Republican), and their age, which we assigned to one of four age bins (18-24, 25-44, 45-64, 65+). 
We use 7-point partisan identification, rather than 3-point (Democrat, Independent, Republican), because strong partisans were over-sampled in the 2018 survey and may differ in important respects from respondents with weaker partisan attachments (see Methods).

To measure exposure (what users saw), we designed our extension to save HTML snapshots of the Google Search results that appeared on participants' screens, and used a custom parser to extract the URLs from those snapshots.
To measure follows from Google Search (exposure followed by an engagement), we collected Google History in 2018, which tracks activity on Google services and is available at \texttt{myactivity.google.com}, and Tab Activity in 2020, a method we developed to improve our measure of engagement by monitoring the active browser tab.
We did not use Google History to measure follows in 2020 because the changes made to that system by Google between 2018 and 2020 were incompatible with our data collection approach.
To measure overall engagement, we collected Browser History, an API that is built into Firefox and Chrome and provides users with an interface for viewing their past website visits.
The third-party data collections---Browser History and Google History---include data prior to the installation of our extension, but our custom data collections---the HTML snapshots and Tab Activity---only collected data while the extension was installed.
Additional details on each dataset are available in Methods.

In 2018, we collected exposure to 105,327 Google Search result pages for 276 participants, follows on 279,680 Google Search results for 262 participants, and overall engagement with 14,677,297 URLs for 333 participants.
In 2020, we collected 
exposure to 227,659 Google Search result pages for 459 participants, follows on 212,796 Google Search results for 459 participants, and overall engagement with 31,202,830 URLs for 688 participants.
Compared to other popular search engines, like Bing and Yahoo, a majority of participants' web searches were conducted on Google Search in both waves (74.2\% in 2018, 70.5\% in 2020; see \autoref{tab:search_engine_usage}).

\subsection{Measuring Web Domains and Search Queries}

Following past work~\citep{bakshy2015exposure,grinberg2019fake,guess2020exposure,peterson2021partisan}, we focus on URLs from news domains, which we identified using a compendium of four lists used in recent research~\citep{grinberg2019fake, yin2018local, newsguard2021rating, bakshy2015exposure} (see Methods).
To classify news domains as unreliable we used a combination of two lists.
The first is from peer-reviewed work that classified news domains as unreliable based on a manual review of their editorial practices~\citep{grinberg2019fake}.
The second was obtained from NewsGuard, an independent organization that employs journalists and editors to review and rate news domains based on nine journalistic criteria~\citep{newsguard2021rating}, and has also been used in recent peer-reviewed work~\cite{bhadani2022political}.

To score the partisanship of news domains we used a measure of partisan audience bias derived from the differential sharing patterns of Democrats and Republicans in a large, virtual panel of Twitter users~\citep{robertson2018auditinga}. 
These scores range from -1 to 1, with a score of -1 indicating that a domain was shared only by Democrats, a score of 1 that it was shared only by Republicans in the Twitter panel, and a score of 0 that an equal proportion of Democrats and Republicans shared it on Twitter. 
For the Google Search results we collected, we aggregated partisanship scores at the Search Engine Result Page (SERP) level by applying a rank-weighted average~\cite{robertson2018auditinga} to the news domains appearing on each SERP.
This measure helps to account for the impact of user attention by placing more weight on the scores of domains that appear near the top of the search rankings, which often receive a disproportionate amount of clicks~\cite{epstein2015search,pan2007google,kulshrestha2019search}.
Additional details for each domain metric are available in Methods.

Unlike the news feeds produced by social media platforms, Google Search result pages are produced through a more active information seeking process that depends on a user-selected search query~\cite{mustafaraj2020case,vanhoof2022searching}.
We therefore measured the content of participants' queries using pivoted text scaling~\cite{hobbs2019text}, 
a form of principal components analysis performed on a truncated word co-occurrence matrix to identify orthogonal latent dimensions that explain decreasing shares of variation in the co-occurrence of commonly-used words. 
This unsupervised approach 
is useful for our application because it does not rely on external sources (such as dictionary-based approaches), nor does it use additional user-level information (such as partisan identities) in the estimation stage that could risk introducing collinearity into subsequent modeling.
It is also appropriate for our use-case of characterizing variation in search queries as it was designed specifically for short documents, where other unsupervised methods such as topic models are less efficient. 
Additional details on this approach 
are available in Methods.

\begin{figure}[!t]
    \centering
    \includegraphics[width=1\textwidth]{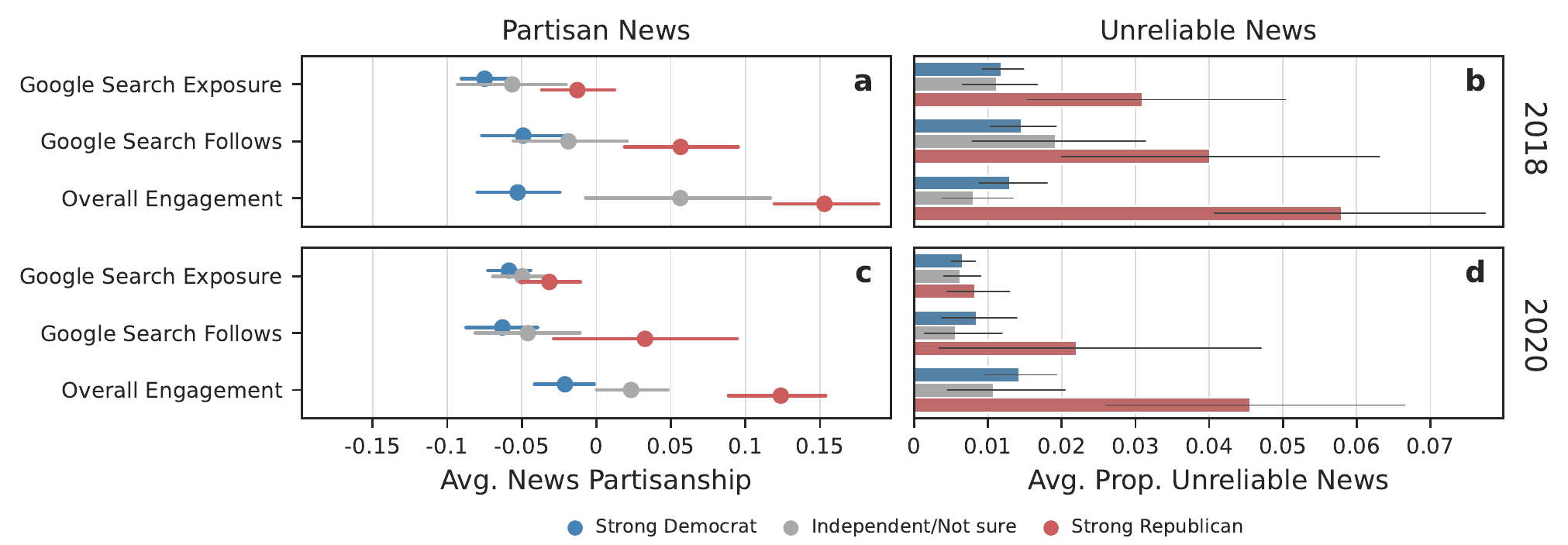}
    \caption{
    \textbf{Strong partisans are exposed to similar rates of partisan and unreliable news, but strong Republicans asymmetrically follow and engage with such news.}
    Average exposure, follows, and overall engagement with partisan (\textbf{a, c}) and unreliable news (\textbf{b, d}) by study wave and 7-point PID clustered at the participant-level.
    For visual clarity, we show only a subset of partisan identity groups here (all groups are shown in Supplementary Information \autoref{fig:partisan_and_unreliable_news_all_groups}).
    The smaller differences in participants' exposure to partisan and unreliable news on Google Search aligns with prior work that found substantial homogeneity among the news domains that Google returns for controlled query sets, which tend to include mainstream national sources that have slightly left-of-zero scores in the metric we used~\cite{trielli2019search,kawakami2020media,fischer2020auditing}.
    Because a score of zero does not imply neutrality~\cite{robertson2018auditinga}, left-of-zero scores do not imply a left-leaning bias (see Methods).
    Results from bivariate tests of differences by partisan identity are available in Supplementary Information \autoref{tab:kruskal_tests}.
    Error bars indicate 95\% confidence intervals (CI).
    }
    \label{fig:partisan_and_unreliable_news}
\end{figure}

\subsection{Results}

In \autoref{fig:partisan_and_unreliable_news}, we compare the average news partisanship of participants' (1) exposure via Google Search results, (2) follows on Google Search results, and (3) overall engagement.
For both study waves, we found that the partisan gap---the difference in news partisanship between the average strong Republican and the average strong Democrat---was small for exposure (0.062 in 2018, 0.037 in 2020), larger for follows (0.106 in 2018, 0.096 in 2020), and largest for overall engagement (0.206 in 2018, 0.134 in 2020).
As in past work, we contextualize these partisan gaps by noting comparable gaps in scores between popular news outlets~\cite{flaxman2016filter}.
For exposure, the partisan gap was comparable to the gap in scores between MSNBC (-0.624) and Mother Jones (-0.697)
for 2018, and the gap between The Washington Post (-0.234) and The New York Times (-0.260) for 2020.
For follows, the partisan gap for both years was comparable to the gap between Newsmax (0.688) and InfoWars (0.782).
Last, for overall engagement, the partisan gap was comparable to the gap between Salon (-0.593) and Jacobin (-0.803) for 2018, and comparable to the gap between Fox News (0.608) and Breitbart (0.742) for 2020.
These results suggest that Google Search's algorithmic curation exposes users to less identity-congruent news than they choose to engage with.

Examining the average proportion of unreliable news participants interacted with by political affiliation, we found a pattern across exposure, follows, and overall engagement that was similar to the one observed for partisan news (\autoref{fig:partisan_and_unreliable_news}).
For the average participant, unreliable news was the least prevalent in their exposure (2.05\% in 2018, 0.72\% in 2020), and more prevalent in both their follows (2.36\% in 2018, 2.03\% in 2020) and their overall engagement (3.03\% in 2018, 1.86\% in 2020).
For exposure, this percentage represents the fraction of SERPs that contained at least one unreliable news domain (URL-level percentages are available in Supplementary Information \autoref{tab:news}).
For engagement, the prevalence of unreliable news was asymmetric across partisan identities, with strong Republicans generally engaging with or following more unreliable news than strong Democrats, but not being exposed to more in their Google Search results.
These results suggest that Google Search's algorithmic curation can be a conduit for exposure to unreliable news, but not to the degree that users' engagement choices are, especially among strong Republicans.

\begin{figure}[!t]
    \centering
    \includegraphics[width=1\textwidth]{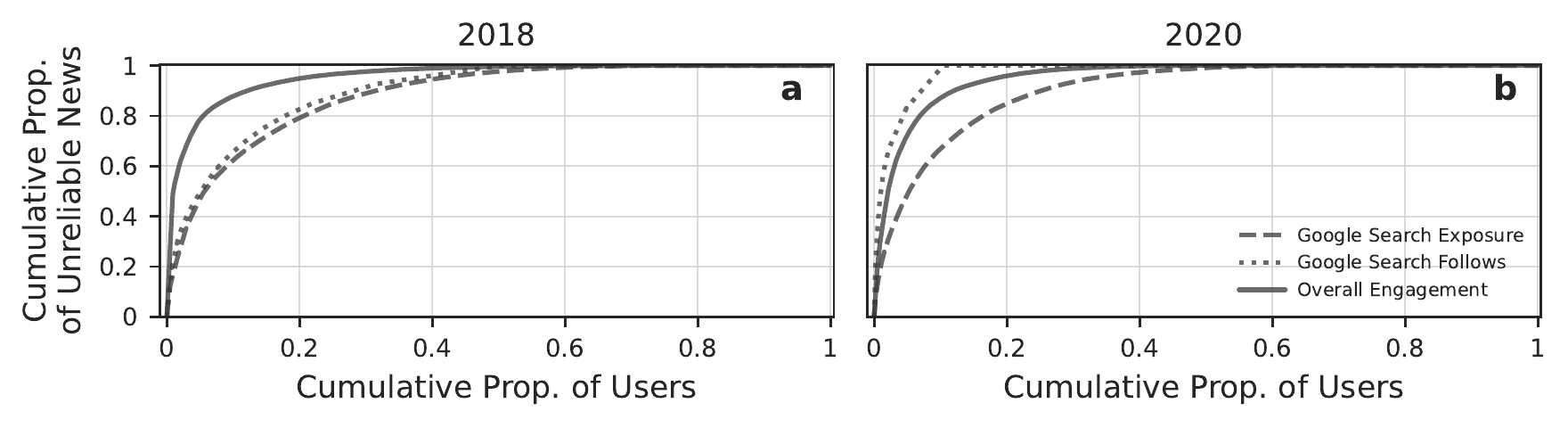}
    \caption{
    \textbf{Exposure to unreliable news is less concentrated among a small number of participants than follows or overall engagement.}
    Each line shows the cumulative proportion of participants (x-axis) that account for the cumulative proportion of unreliable news seen by all participants (y-axis) within each data type (see legend) and each study wave (\textbf{a,b}).
    }
    \label{fig:unreliable_news_concentration}
\end{figure}

Overall, we found that unreliable news was generally uncommon and concentrated among a small number of participants (\autoref{fig:unreliable_news_concentration}).
Using the percentage of participants that accounted for 90\% of all exposures, follows, or engagements as a measure of concentration, we found a pattern similar to the user-level averages.
Specifically, we found that unreliable news was the least concentrated among a small number of participants for exposure, with 31.3\% of participants accounting for 90\% of all unreliable news exposures in 2018 (25.1\% in 2020), and more concentrated in both follows (28.2\% in 2018, 7.2\% in 2020) and overall engagement (12.0\% in 2018, 11.9\% in 2020).
Gini coefficients calculated for each study wave and dataset suggest a similar pattern (see Supplementary Information~\ref{sec:descriptives_supp}).
These findings suggest that interactions with unreliable news are driven by a relatively small number of individuals, especially for follows from Google Search and overall engagement.

\begin{figure}[!t]
    \centering
    \includegraphics[width=1\textwidth]{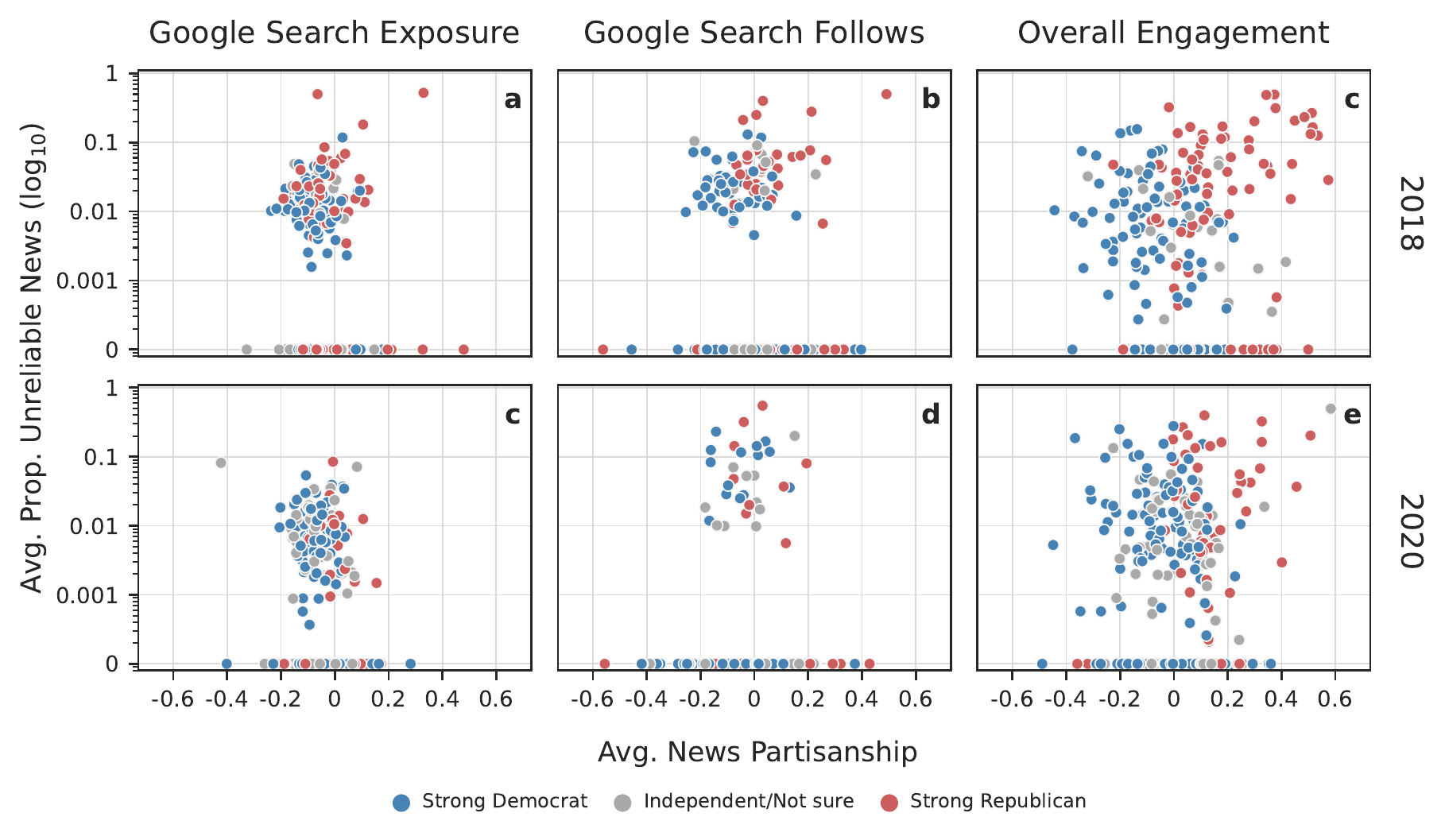}
    \caption{
    \textbf{Partisans who engage with more identity-congruent news also tend to engage with more unreliable news.}
    The relationship between partisan and unreliable news for participants exposure on Google Search (\textbf{a, d}), follows from Google Search (\textbf{b, e}), and overall engagement (\textbf{c, f}).
    For visual clarity, we show only a subset of partisan identity groups here (all groups are shown in Supplementary Information \autoref{fig:partisan_and_unreliable_news_all_groups}).
    These plots highlight how the relationship between partisan and unreliable news varies across data types, and within data types when taking partisan identity into account.
    }
    \label{fig:partisan_by_unreliable_news}
\end{figure}

In \autoref{fig:partisan_by_unreliable_news} we examine the relationship between partisan and unreliable news by comparing participant-level averages across data types and study waves.
We found a significant positive correlation between overall engagement with partisan and unreliable news among strong Republicans in 2018 ($\rho = 0.336; P < 0.001$), and a significant negative correlation among strong Democrats in both 2018 ($\rho = -0.380; P < 0.01$) and 2020 ($\rho = -0.237; P < 0.01$).
In contrast, we found no significant correlations between partisan and unreliable news exposure or follows.
Additional group and overall correlations are available in Supplementary Information \autoref{tab:spearman_tests}.
These findings suggest that, on average, partisans who choose to engage with more ideologically-congruent news also have a greater proclivity for engaging with unreliable news, but this association does not carry over into their Google Search results or follows.

In a series of multivariate regressions that accounted for participants' demographic characteristics and query text features, we found inconsistent relationships between partisan identity and exposure to partisan news on Google Search.
For example, although we found small but significant differences between strong Republicans and independents' exposure to partisan news on Google Search in 2020, 
these differences were not significant when accounting for participants' queries (\autoref{tab:bias_search_2020}).
In both study waves, participants' query text features exhibited significant associations with the rank-weighted average partisanship scores 
of the domains they were shown 
in their search results (Tables \ref{tab:bias_search_2018} and \ref{tab:bias_search_2020}).
In 2018, all age groups saw significantly more right-leaning news on Google Search than participants aged 18-24 on average, with participants aged 65+ exhibiting the largest such differences.
This trend also appeared in 2020, though only the oldest (65+) age group had significant differences that were robust to the inclusion of query text features (Model 5 in Table \ref{tab:bias_search_2020}).
These results suggest that, to the extent there are differences in exposure to partisan news via Google Search results, they are small, inconsistent, and at least partially attributable to differences in users' queries and demographic characteristics.

In contrast, our regressions for participants' follows from Google Search show that strong Republicans followed significantly more right-leaning sources than independents (by 0.09 in 2018, and 0.06 points in 2020) after accounting for query text features.
However, strong Democrats did not exhibit 
similar significant differences in any model specification for either study wave. 
As with exposure, age was significantly associated with follows to partisan news across model specifications (\autoref{tab:bias_follows_2018} and \ref{tab:bias_follows_2020}), with participants in the 45-64 and 65+ age bins following the most right-leaning news.
These results show that, even when accounting for the contents of their search queries, right-leaning partisans are more likely to follow right-leaning sources from Google Search than independents.

For overall engagement with partisan news, we found that strong Democrats and strong Republicans consistently engaged with significantly more identity-congruent news than independents in both study waves, but this association was less consistent for other partisan identities. 
When accounting for all control variables, strong Democrats chose to engage with more left-leaning news (-0.10 in 2018, -0.03 in 2020) and strong Republicans chose to engage with more right-leaning news (0.10 in 2018, 0.08 in 2020) than independents did.
These associations hold across model specifications (\autoref{tab:bias_browse_2018} and \ref{tab:bias_browse_2020}), and age was again significantly associated with more right-leaning news, with the greatest association for those aged 65+ in both study waves.
We did not include query text features in these regressions, as overall engagement does not directly depend on participants' Google Search queries.

Among our 2018 sample, we found that strong Republicans were exposed to more unreliable news in their Google Search results than independents, but this association was not robust to the inclusion of demographic characteristics 
and query text features (Models 4 and 5; \autoref{tab:unreliable_search_2018}).
In contrast, we found no significant relationship between partisan identity and exposure to unreliable news in any model specification for 2020 (\autoref{tab:unreliable_search_2020}).
We also found no differences in follows to unreliable news by partisan identity that were robust to the inclusion of query text features in 2018 (\autoref{tab:unreliable_follows_2018}), and in 2020 the only robust relationship was for participants who identified as Strong Republicans (\autoref{tab:unreliable_follows_2020}).
Similarly, the only systematic relationships between age group and unreliable news follows that was robust to the inclusion of additional demographic characteristics and query text features was the 45-64 year-old group following significantly more links to such news than 18-24 year-olds in 2020.

For overall engagement with unreliable news, we found that strong Republicans, but not strong Democrats, engaged with significantly more news from unreliable sources than independents.
These results are robust to model specifications and consistent across both study waves (Tables \ref{tab:unreliable_browse_2018} and \ref{tab:unreliable_browse_2020}).
Older participants also engaged with significantly more unreliable news in both waves, but in 2018 this relationship was not robust to the inclusion of additional demographic characteristics (Model 4 in Table  \ref{tab:unreliable_browse_2018}).

\subsection{Discussion}

The two waves of our study replicate the same finding: 
engagement outweighs exposure to partisan and unreliable news within Google Search, and the small differences we observe in exposure are largely explained by query selection.
This pattern is consistent across data collected during two distinct time periods, each with a different social, political, and technological context. 
For concerns related to filter bubbles and echo chambers, our results highlight the role of user choice, rather than algorithmic curation, in driving such effects. 
These findings add to the limited number of studies examining ecological exposure~\cite{huszar2022algorithmic}, align with prior work on Facebook's News Feed~\citep{bakshy2015exposure}, and are consistent with studies that have found engagement with unreliable or problematic content to be rare and concentrated among a small number of individuals~\citep{grinberg2019fake,hosseinmardi2021examining,guess2019less}.
Future work should examine the mechanisms driving the behavioral patterns we observed by taking a broader view of the information systems involved, both on- and offline~\citep{allen2020evaluating,muise2022quantifying,wagner2021measuring,zuckerman2021why}.

These findings do not necessarily imply that the design of Google Search's algorithms is normatively unproblematic. 
In some cases our participants were exposed to highly-partisan and unreliable news on Google Search, and past work has demonstrated that even a limited number of such exposures can have substantial negative impacts~\citep{noble2018algorithms,epstein2015search,guess2021consequences}.
However, determining the circumstances under which such content should be shown or omitted is complex, especially when considering who should make such decisions~\citep{introna2000shaping}.
As 
data sharing agreements between academia and industry continue to stall,
our approach provides an independent avenue 
toward algorithmic accountability~\cite{diakopoulos2015algorithmic} by collecting data directly from real users under ecological conditions.

One limitation of our study is that 
we rely on domain-level metrics to identify and score partisan and unreliable news sites, which limits our ability to detect differences that occur within a given domain (\eg unreliable news from the New York Times).
Another limitation is that we only collected data from desktop computers, whereas mobile devices increasingly serve as an avenue to online news~\citep{walker2019americans}.
Last, we only collected the first page of search results because the majority of searchers do not navigate beyond that~\cite{pan2007google,epstein2015search}, and collecting results past the first page could provide additional context on Google's algorithmic curation.
Despite these limitations, our study provides an ecologically-valid look at 
exposure and engagement within Google Search, advances methods for collecting such data on any platform, and aligns with the past work by pointing toward the importance of examining platforms in the context of their users' broader online behavioral patterns~\citep{rahwan2019machine,bakshy2015exposure}.

\printbibliography

\newpage
\input{methods}

\newpage
\include{supplementary}

\end{document}

%% file: abstract.tex
\begin{abstract}

If popular online platforms systematically expose their users to partisan and unreliable news, they could potentially contribute to societal issues like rising political polarization.
This concern is central to the ``echo chamber'' and ``filter bubble'' debates, which critique the roles that user choice and algorithmic curation play in guiding users to different online information sources.
These roles can be measured in terms of
\textit{exposure}, defined as the URLs seen while using an online platform, and \textit{engagement}, defined as the URLs selected while on that platform or browsing the web more generally.
However, due to the challenges of obtaining ecologically valid exposure data---what real users saw during their regular platform use---studies in this vein often only examine engagement data, or estimate exposure via simulated behavior or inference.
Despite their centrality to the contemporary information ecosystem, few such studies have focused on web search, and even fewer have examined both exposure and engagement on any platform.
To address these gaps, we conducted a two-wave study pairing surveys with ecologically valid measures of both exposure and engagement on Google Search during the 2018 and 2020 U.S. elections.
In both waves, we found that participants' partisan identification had a small and inconsistent relationship with the amount of partisan and unreliable news they were exposed to on Google Search, a more consistent relationship with the search results they chose to follow, and the most consistent relationship with their overall engagement.
That is, compared to the news sources our participants were exposed to on Google Search, we found more identity-congruent and unreliable news sources in their engagement choices, both within Google Search and overall.
These results suggest that exposure and engagement with partisan or unreliable news on Google Search are not primarily driven by algorithmic curation, but by users' own choices.

\end{abstract}

%% file: methods.tex
\section{Methods}

\subsection{Opinion Surveys and Participant Samples}

The self-reported data used in this study was based on two multi-wave public opinion surveys. 
In each of them, respondents were asked to install a browser extension that would monitor their Google Search results and various aspects of their online activity. 
The protocol and informed consent language we used while recruiting subjects was transparent about what the extension collected, and both studies were approved by the IRB at Northeastern University (\#18-10-03 for 2018, and \#20-03-04 for 2020).

The first survey was fielded between October 18 and October 24 of 2018, and participants were recruited through YouGov with an oversample of strong partisans. 
The second was fielded between June 2020 and January 2021, with participants recruited by PureSpectrum as part of the ``COVID States Project'' (see \url{https://covidstates.org}).
This second survey used quota sampling based on state-by-state benchmarks for race, gender, and age, but a non-random subset of these respondents opted-in to installing our browser extension.

In both surveys, participants self-reported a seven-point partisan identification (7-point PID; strong Democrat, not very strong Democrat, lean Democrat, Independent, lean Republican, not very strong Republican, strong Republican), their age, which we binned (18--24, 25--44, 45--64, 65+), and several other demographic variables (race, gender, education).
We use a seven-point partisan scale, rather than a three-point scale (Democrat, Independent, Republican) for two reasons. 
First, binning all Republicans and Democrats into a single category would obscure our intentional oversampling of strong partisans in 2018.
Second, recent work in political science has found that independents who lean toward a political party tend to behave in a more partisan manner than weak partisan identifiers \citep{klar2016independent}, further complicating the use of a three-category measure.

Due to budget constraints around sample size, we anticipated a relatively small sample size in 2018, and intentionally over-sampled on strong partisans.
As a result, approximately 36\% of our participants identified as Strong Democrats and 27\% identified as Strong Republicans (\autoref{tab:demographics2018}). 
Ensuring that we had a significant number of such people in our sample was key to our investigation because engagement with partisan and unreliable news has been shown to be uncommon and concentrated among a relatively small number of predominantly partisan individuals~\citep{grinberg2019fake,guess2019less,guess2020exposure}.
In 2020, respondents were selected via quota sampling based on population benchmarks in U.S states for several categories (race, gender, and age), and 27\% of identified as Strong Democrats while 13\% identified as Strong Republicans (\autoref{tab:demographics2020}).

\subsection{Exposure and Engagement Definitions}

We broadly define \textit{exposure} as a measure of the content appearing on a user's screen, including links, text, and media. 
In this study, we specifically measure exposure to links that appeared on a participant's screen after they conducted a Google search. 
Our approach and definition for exposure are related to concepts including ``involuntary exposure''~\cite{cardenal2019digital}, ``passive exposure''~\cite{allen2020evaluating}, ``distributed access''~\cite{fletcher2021more}, and ``incidental exposure''~\cite{nanz2022democratic}, but most closely align with \citet{bakshy2015exposure}, who defined an ``exposure'' as a case ``in which a link to the content appears on screen in an individual's [Facebook] NewsFeed.''
Our approach for capturing exposure is also analogous to that used by the Screenome project, which preserves what users saw through the ``collection of high-density sequences of screenshots''~\cite{reeves2021screenomics}.
However, instead of periodically collecting image screenshots, we took HTML snapshots when participants visited a given website, which has the advantage of enabling programmatic extraction of the URLs that appeared on the screen. 

We broadly define \textit{engagement} as a measure of the actions taken by a user in their web browser, including URLs clicked, URLs typed or copy-pasted, and browser tab switches. 
Here we measured engagement using several digital trace datasets that we collected through existing and custom data collection tools. 
Our use of engagement aligns with terms from prior work including ``visits,'' ``clicks,'' and ``consumption''~\cite{allen2020evaluating,bakshy2015exposure,guess2021almost}, ``direct access''~\cite{fletcher2021more}, and ``choice'' or ``voluntary exposure''~\cite{cardenal2019digital}.
While these terms share an emphasis on user choice, ``engagement'' expands the scope of behaviors considered and provides a high-level term for digital trace logs with varying granularity but common measurement aims.

Using the above definitions, we broadly define a \textit{follow} as an engagement conditional on an exposure.  
That is, a follow is an instance in which a person is exposed to search results, a newsfeed, or recommendations during a visit to an online platform, and subsequently visits a URL within the next minute.
This approach is in line with prior work on identifying follows, which have previously been called ``referrals''~\cite{allen2020evaluating}, and their preceding platforms visits ``referrers''~\cite{guess2020exposure}.
For example, a visit to BBC.com might be attributed to Facebook if a visit to Facebook preceded that visit, and occurred within a short (e.g. 1 min) time threshold~\cite{allen2020evaluating,stier2021post,flaxman2016filter,fletcher2021more}. 
Another approach involves examining news article visits and checking if a platform visit occurred within the past three URLs visited and prior 30 seconds~\cite{guess2020exposure}.
We use follows instead of referrals here because the term ``referral'' collides with the HTTP ``referer'' header field (which we did not collect), which may contain information about the previously visited website, but has evolved over time, is omitted under certain circumstances, and can be modified by both the user and the ``referring'' platform.

\subsection{Exposure and Engagement Datasets}

We built custom browser extensions in 2018 and 2020 to collect several types of digital trace data that would enable us to compare the URLs participants saw on Google Search (exposure), the results they followed from Google Search results (follows), and the URLs they engaged with on the web in general (engagement).
In terms of exposure, the extension passively---by monitoring the participant's web browsing activity---collected snapshots of their \textit{Google Search} results. 
For overall engagement, the extension actively---through automated periodic requests---collected each participant's \textit{Browser History}.
To measure follows in 2018, the extension also actively collected their \textit{Google History}, which consists of clicks on Google Search and related Google services, but did not collect the same data in 2020 due to changes in how that website functioned.
To measure follows in 2020, we introduced a passive measure of engagement that tracked the participant's active browser tab, which we call \textit{Tab Activity}.

In both waves, we filtered out participants who did not meet certain web activity levels. 
Specifically, we filtered out participants whose participation window---the duration between their first and last observed behavior---was less than 10 days, which suggests that they installed and then quickly uninstalled the extension.
For the 2018 datasets, this led to filtering out the Browser History of 7 participants, the Google History for 5 participants, and Google Search results for 30 participants.
Using the same rules for our 2020 datasets, we filtered out Browser History for 32 participants, Tab Activity for 234 participants, and Google Search results for 182 participants.
These filtering decisions did not substantially remove more participants from one 7-point PID category than another.
In \autoref{tab:counts} we report overall data counts, including the number of searches conducted and the number of URLs clicked.

In accordance with IRB guidelines, full consent was obtained and participants were compensated for installing the extension.
Participants were informed that they could uninstall the browser extension at any time, and at the end of the data collection period, the extension automatically uninstalled itself.
Additional details on our IRB study procedure are available in Supplementary Information.

\subsubsection{Google Search Results}

For each search our participants conducted while our browser extension was installed, we saved an HTML snapshot of the corresponding Search Engine Results Page (SERP) that they were exposed to on Google Search.
We only collected the first SERP because most search traffic goes to top-ranked results, and most users do not go past the first page of results~\citep{pan2007google}.
The extension did not collect snapshots of Google Searches in incognito (private) browser windows.
We then parsed these SERPs, extracting detailed information from each result, including a URL (if present) and a classification of the result format (\eg News, Twitter, or Knowledge results).
The SERP parser we used is available at: \url{https://github.com/gitronald/websearcher}.

\subsubsection{Google History}

For the Google History dataset, we collected participants' Google account activity, which provided us with a log of their clicks on Google SERPs, as well as their activity on Google News and other Google products.
We collected these data by identifying an unofficial API endpoint and then incrementally collecting data through requests periodically sent by the browser extension.
These requests occurred every two weeks throughout the study.
The data types we collected can be seen by going to the My Activity page (\url{https://myactivity.google.com}) while logged into a Google account.
To ensure that we could access that page, we asked participants to remain logged in for the duration of the study, and sent them reminders to log in if they were not when the extension periodically attempted to collect the latest data.
We identified and removed consecutive visits to the same web page that occurred within one second, keeping only the first instance.
Consecutive visits to the same web page are often present in such data, and can occur for a variety of reasons, such as refreshing a stalled page, or website-specific idiosyncrasies in page loading.

\subsubsection{Tab Activity}

In the 2020 extension, we designed Tab Activity to log changes in the active browser tab. 
This overcomes limitations with current approaches to collecting engagement data.
For example, it is often unclear exactly how data from proprietary sources are obtained, logged, or cleaned. 
In contrast, Browser History---which is maintained by Chrome and Firefox---has public documentation which suggests it provides a log of the first time a webpage loads in the browser, but does not account for changes in the active browser tab. 
By passively monitoring the active browser tab, Tab Activity provides a more direct and detailed measure of user attention than other engagement datasets---such as Browser History---by accounting for the common practice of tabbed browsing. 
We removed duplicates from Tab Activity using the same method as we did in Google History, removing sequential duplicates that occurred within a one second interval of each other.

\subsubsection{Browser History}

We collected Browser History by accessing an API\footnote{\scriptsize\url{https://developer.mozilla.org/en-US/docs/Mozilla/Add-ons/WebExtensions/API/history}} that is built into Chrome and Firefox.
In both data collection waves, the extension collected data every two weeks for the duration of the study. 
An important difference between the Browser History provided by Chrome and Firefox is that Chrome allows you to access only the past 3 months of Browser History, whereas Firefox allows you to continue saving history regardless of time passed, but enforces a maximum page limit.\footnote{\scriptsize\url{https://support.mozilla.org/en-US/questions/1059856}}
In 2018, we collected a pre-aggregated version of Browser History (HistoryItems\footnote{\scriptsize\url{https://developer.mozilla.org/en-US/docs/Mozilla/Add-ons/WebExtensions/API/history/HistoryItem}}),
while in 2020 we collected a higher resolution data that captured each website visit (VisitItems\footnote{\scriptsize\url{https://developer.mozilla.org/en-US/docs/Mozilla/Add-ons/WebExtensions/API/history/VisitItem}}).
For all analyses, we combined visits and typed counts into a total count.
Due to the API-level aggregation of the 2018 Browser History data, we did not apply duplicate correction to it.
However, for the 2020 data we identified and removed consecutive visits to the same web page that occurred within one second, as we did with Google History and Tab Activity.

\subsection{Scoring and Classifying News Domains}

We classified and scored online news at the domain-level.
That is, for each URL in our dataset, we extracted the second-level domain name (\eg https://www.cnn.com $\rightarrow$ cnn.com).
This enabled us to merge our data with several datasets that contain domain-level scores and classifications for news, partisanship, and reliability.

\subsubsection{News Classifications}

To identify visits to news domains, we compiled four datasets containing classifications of web domains as ``news.''
After cleaning, the union of these datasets gave us 11,902 unique domains coded as news.
These datasets include:
\begin{packed_enumerate}
    \item 488 domains identified as ``hard news'' by Bakshy et al. (2015)~\cite{bakshy2015exposure},
    \item 1,250 domains manually identified as news by Grinberg et al. (2019)~\cite{grinberg2019fake},
    \item 6,288 domains aggregated from local news listings by Yin (2018)~\cite{yin2018local}, 
    \item 6,117 domains identified as news by NewsGuard (2021)~\cite{newsguard2021rating}
\end{packed_enumerate}

Bakshy et al. (2015) used the URLs and associated text snippets shared by Facebook users who self-reported a partisan identity to classify ``hard news'' domains, and publicly released the 500 most shared domains~\cite{bakshy2015exposure}.
As in prior work using these classifications~\cite{guess2021almost}, we exclude their classifications of five platforms\footnote{We excluded platform domains: youtube.com, m.youtube.com, amazon.com, twitter.com, and vimeo.com} as news.
We also excluded a satire site (theonion.com) and Wikipedia\footnote{\scriptsize\url{https://en.wikipedia.org/wiki/Wikipedia:Wikipedia_is_not_a_newspaper}} as news sites, and merged five entries that had a duplicate including a ``www.'' prefix, which typically direct to the same homepage, leaving us with 488 domains.
The list used by Grinberg et al. (2019) was manually curated and contains a list of 1,250 domains coded as news based on editorial practices~ \cite{grinberg2019fake}.

Yin (2018) provides a list of 6,290 domains associated with state newspapers, TV stations, and magazines, aggregated from several sources, including the United States Newspaper Listing~\cite{yin2018local}.
We removed a string value (``Alaskan Broadcast Television'') and the coding of myspace.com (a social media platform) as news domains from the current version of this dataset.
NewsGuard is an independent organization that ``employs a team of trained journalists and experienced editors to review and rate news and information websites based on nine journalistic criteria.''
To maintain consistency with the other lists, we classified all domains covered by NewsGuard as news except those labelled as a satire site (``not a real news website'') or a platform (``primarily hosts user-generated content that it does not vet''), as neither receives news ratings based on the nine journalistic criteria~\citep{newsguard2021rating}.

We measured the proportion of news present in Google's SERPs in two ways.
At the URL-level, we used the mean proportion of links on a SERP that led to news domains.
At the SERP-level, we used the mean proportion of SERPs containing at least one news domain.
Counts and averages for both measures are available in \autoref{tab:counts}, \autoref{tab:user_level}, and \autoref{tab:news}.

\subsubsection{Partisan News Scores} 

To quantify participants' exposure and engagement with partisan news, we used the \textit{partisan audience bias} scores developed in prior work~\citep{robertson2018auditinga}.
These scores were made using the domain sharing patterns of a large virtual panel of Twitter users who had been linked to US voter registration records.
More specifically, each domain was scored by comparing the relative proportion of registered Democrats and Republicans who shared it on Twitter.
The resulting scores range from -1 (shared only by Democrats) to 1 (shared only by Republicans). 
On this scale, a domain score of 0 does not mean neutral or unbiased, only that an equal proportion of Democrats and Republicans in the virtual panel shared it.

These scores have been used in several recent examinations of engagement with partisan news~\cite{garimella2021political,muise2022quantifying,wojcieszak2021avenues}, provide coverage for more domains overall ($N=19,022$)\footnote{We resolved a small number of duplicates by keeping the item whose score was based on a greater number of accounts that shared the domain, leaving us with scores for 19,014 domains.} 
and news domains specifically ($N=2,584$, covering 21.7\% of the 11,902 domains we classified as news) than other domain-level partisanship metrics, and are strongly correlated ($r = 96^{***}$) with the most widely used alternative~\citep{bakshy2015exposure}.
However, similar to other domain-level metrics~\cite{bakshy2015exposure}, these scores are limited in terms of the context of the shares they're based on: a user may share a news article from a domain because they are denouncing it, not because they support it.

We aggregated partisanship scores to the user-level in slightly different ways for our exposure and engagement data.
For exposure to Google Search results, we calculated the partisanship of each SERP using a weighted average that takes ranking into account~\cite{robertson2018auditinga} and places more weight to the partisan audience scores of domains appearing near the top of the search rankings.
This helps to account for the additional attention received by highly ranked search results~\citep{robertson2018auditinga, kulshrestha2019search}, which is partly due to position effects that have been long studied in the psychological sciences~\cite{ebbinghaus1913memory} and have been found to affect search engine user behavior~\cite{epstein2015search,pan2007google}.

Weighted average bias $B^{w}$ is calculated by finding the bias $B$ until reaching rank $r$ for a query $q$, given each domain score $s_i$ : $B(q, r) = \sum_{i=1}^{r} s_i / r$, and then taking the normalized sum of this over all ranks: $B^{w}(q, r) = \sum_{i=1}^{r} B(q, i) / r$.
We then used those rank-weighted averages to calculate the average news partisanship of each participant's search results.
For our engagement data, we calculated a user-level average by taking the mean score of all the news domains they visited during our study.
To handle the aggregated Browser History data we collected in 2018, we multiplied each domain's score by the number of visits to that domain, and then divided the sum by the total number of news visits.
We report group differences throughout the paper using these user-level scores.

\subsubsection{Unreliable News Classifications} 

We classified 2,962 web domains as unreliable if they appeared in either of two carefully constructed lists~\cite{newsguard2021rating, grinberg2019fake}.
First, NewsGuard (introduced in the News Classifications section) tracks over 6,000 news websites for information quality and assigns each a score from 0 (unreliable) to 100 (reliable)~\cite{newsguard2021rating}. 
We used the threshold defined by NewsGuard to classify news domains, and labelled each of the 2,534 domains with a score of less than 60 as unreliable. 
Second, Grinberg et al. (2019) investigated fake news sharing on Twitter and manually assembled their list of unreliable news domains~\cite{grinberg2019fake}.
These classifications were made by examining fact checkers evaluations of stories produced by various domains, and defined fake news as content that has the form of standard media, but not the intent or processes to produce accurate content. 
The colors they used to code domains include: Black if they contained ``almost exclusively fabricated stories,'' Red if they ``spread falsehoods that clearly reflected a flawed editorial process,'' and Orange in ``cases where annotators were less certain that the falsehood stemmed from a systematically flawed process.''
As in the original paper, we consider all 490 domains coded as Black, Red, or Orange to be unreliable news domains.

\subsection{Descriptive Results}

To assess overall search engine use, we identified URLs leading to popular web search engines by filtering for known domains (\eg \texttt{google.com}) and a URL path indicating a page of search results (\eg \texttt{google.com/search}), which excludes visits to each search engine's homepage as in prior work~\cite{allen2020evaluating}. 
We found that Google handled a substantial majority of our participants' web searches in both 2018 (74.2\%) and 2020 (68.6\%), and Bing handled the second most (21.7\% in 2018; 30.5\% in 2020). 
This majority use of Google Search is in line with industry estimates of Google's desktop market share~\cite{fishkin2018new,statcounter2020desktop}), provides support for our focus on this online platform, and we provide additional details on overall search engine use in Supplementary Information Table~\ref{tab:search_engine_usage}.

In Tables \ref{tab:counts}, \ref{tab:user_level} and \ref{tab:news} we report key counts and participant-level averages across data types and study waves.
For the average participant in either study wave, we found a substantially greater proportion of news in the Google Search results that they were exposed to (14.3\% for 2018, 14.7\% for 2020), than we did in amount of news they chose to engage with overall (7.1\% for 2018, 3.2\% for 2020).
For partisan and unreliable news, we also checked for statistically significant differences by partisan identity and age group using the Kruskal-Wallis (KW) test, a nonparametric test of differences among three or more groups. 
We used this nonparametric test because of the heterogeneity we observed in the distribution of user-level averages.
Results for each KW test are available in \autoref{tab:kruskal_tests}.
Similarly, we evaluated the relationship between partisan and unreliable news using Spearman's rank correlation coefficient ($\rho$), another nonparametric test, and these results are available in \autoref{tab:spearman_tests}.

\subsection{Search Query Analysis}

To quantify the content of participants' search queries, we used pivoted text scaling~\cite{hobbs2019text}. 
Pivoted text scaling is a form of principal components analysis performed on a truncated word co-occurrence matrix, identifying orthogonal latent dimensions that explain decreasing shares of variation in the co-occurrence of commonly-used words. 
Each word is then given a score along each dimension, and each document can then be scored with respect to each dimension based on its average word scores (which we then further aggregate to the participant level). 
We incorporate participants' average document score along each of the first nine dimensions derived from their respective corpora, and include these in the regressions that account for query text (Model 5 in Tables \ref{tab:bias_search_2018}-\ref{tab:bias_follows_2020} and Tables \ref{tab:unreliable_search_2018}-\ref{tab:unreliable_follows_2020}).

This measure is well suited for the purposes of our study for multiple reasons.
First, as it is unsupervised, it does not require the use of external dictionaries of partisan speech developed in other contexts, such as politicians' speech, which may not be applicable for search queries.
Second, it does not rely on additional user-level information, such as their partisan identities, that could introduce collinearity into subsequent regressions that use our quantitative representations of users' query text.
Finally, the method was designed specifically for short documents---such as social media posts or, in our case, search queries---where other unsupervised methods such as topic models are less efficient.

\subsection{Multivariate Regressions}

We ran a series of regressions to estimate the associations between several theoretically-motivated factors and exposure, follows, and engagement with partisan or unreliable news. 

\para{Outcomes and Model Selection.} 
For partisan news, our outcome is the average partisan audience score described above, and we estimate this outcome using ordinary least squares.
For unreliable news, our outcome is the count of either URLs from unreliable sources or the count of search engine result pages that contain at least one unreliable source.
As these count outcomes are over-dispersed---in that their variances are greater than their means---we estimate them using negative binomial regressions.

\para{Independent Variables} 
Drawing on prior research \citep{grinberg2019fake,guess2020exposure,guess2019less} and our descriptive results above, our primary independent variables are age and partisan identity. 
In addition, we consider the following additional adjustment variables for each respondent:
\begin{packed_enumerate}
    \item race (white/non-white),\footnote{For limitations of using binary variables for race and gender see \cite{brown2014political} and \cite{hyde2019future}.}
    \item gender (male/female),
    \item education (college/non-college), 
    \item news interest (high/low),\footnote{Whether the respondent reports high levels of news interest---in the 2018 survey this corresponds to reporting that they follow news and current events ``most of the time'' and, in the 2020 survey, reporting being ``very'' or ``extremely'' interested in US politics and government.}
    \item query text
\end{packed_enumerate}

We iteratively build our models around partisan identity and age because prior research identified them as being associated with exposure and engagement with partisan and unreliable news~\citep{grinberg2019fake,guess2019less,guess2020exposure,guess2021almost}.
As such, we build out each model iteratively with four specifications of independent variables:
\begin{enumerate}
    \setlength{\itemsep}{2pt}
    \setlength{\parskip}{0pt}
    \setlength{\parsep}{0pt}
    \setlength{\topsep}{2pt}
    
    \item partisan identity,\footnote{Represented as a seven-category factor variable with independent taken as the reference category.}
    \item age group
    \item age group and partisan identity,
    \item age group, partisan identity, and demographic controls
\end{enumerate}

For outcomes that directly depend on search results, we include an additional specification that includes query text features:

\begin{enumerate}[resume]
    \setlength{\itemsep}{2pt}
    \setlength{\parskip}{0pt}
    \setlength{\parsep}{0pt}
    \setlength{\topsep}{2pt}
    
    \item partisan identity, age group, demographic controls, and average document score along first nine pivot scale dimensions of query text
\end{enumerate}

We do not include query text features in regressions with outcomes based on users' browser history, as these outcomes do not directly depend on users' search behavior or results.

\para{Regression Results.} Regression results for each outcome (partisan bias and unreliable news in Google Search results, follows from Google Search, and overall engagement) are available in the following Supplementary Information tables:

\noindent Partisanship:
\begin{packed_enumerate}
\item Google Search Results: \autoref{tab:bias_search_2018} [2018] and \autoref{tab:bias_search_2020} [2020]
\item Follows from Search: \autoref{tab:bias_follows_2018} [2018] and \autoref{tab:bias_follows_2020} [2020]
\item Browser History: \autoref{tab:bias_browse_2018} [2018] and \autoref{tab:bias_browse_2020} [2020]
\end{packed_enumerate}

\noindent Unreliable News:
\begin{packed_enumerate}
\item Google Search Results: \autoref{tab:unreliable_search_2018} [2018] and \autoref{tab:unreliable_search_2020} [2020]
\item Follows from Search: \autoref{tab:unreliable_follows_2018} [2018] and \autoref{tab:unreliable_follows_2020} [2020]
\item Browser History: \autoref{tab:unreliable_browse_2018} [2018] and \autoref{tab:unreliable_browse_2020} [2020]
\end{packed_enumerate}

\subsection{Data availability}

Due to privacy concerns and IRB limitations, aggregated data are available upon request, but individual data cannot be shared. The domain scores and classifications we used are available at \url{https://github.com/gitronald/domains}, but the NewsGuard classifications are not included due to their proprietary nature.

\subsection{Code availability}

The data for this study were collected using custom browser extensions written in JavaScript and using the WebExtension framework for cross-browser compatibility. The source code of the extensions used in 2018 and 2020 will be posted upon publication. The parser we used to extract the URLs our participants were exposed to while searching is available at \url{https://github.com/gitronald/WebSearcher}. Analyses were performed with Python v3.10.4, pandas v1.4.3, scipy v1.8.1, Spark v3.1, and R v4.1. 

\para{Acknowledgements} Early versions of this work were presented at the 2019 International Conference on Computational Social Science (IC2S2), the 2019 Conference on Politics and Computational Social Science (PaCCS), and the 2020 annual meeting of the American Political Science Association (APSA). We are grateful to the NYU SMaPP lab and the Stanford Internet Observatory for feedback, and to Muhammad Ahmad Bashir for development on the 2018 extension. This research was supported in part by the Democracy Fund, the William and Flora Hewlett Foundation, and the NSF (IIS-1910064).

\para{Author Contributions}
R.E.R., C.W., and D.L. conceived of the research, K.O., C.W., D.L., and R.E.R. contributed to survey design, R.E.R. built the 2020 data collection instrument, J.G. designed the multivariate regression analysis, R.E.R. and J.G. analyzed the data, and R.E.R. wrote the paper with D.R., J.G., K.O., C.W., and D.L. All authors approved the final manuscript.

\noindent\textbf{Additional information}

\para{Correspondence and requests for materials} should be addressed to ronalder@stanford.edu.

\para{Competing interests} The authors declare no competing interests.

%% file: supplementary.tex
\setcounter{equation}{0}
\setcounter{figure}{0}
\setcounter{table}{0}
\setcounter{page}{1}
\setcounter{secnumdepth}{2} %

\makeatletter
\renewcommand{\theequation}{S\arabic{equation}}
\renewcommand{\thefigure}{S\arabic{figure}}
\renewcommand{\thetable}{S\arabic{table}}
\renewcommand{\thepage}{S\arabic{page}}
\renewcommand{\thesection}{S\arabic{section}}

\begin{center}
\fontsize{18pt}{18pt}\selectfont\bfseries
Supplementary Information for \\ ``Engagement Outweighs Exposure to Partisan and Unreliable News within Google Search''
\end{center}

\section{Supplementary Methods}

Here we provide supplementary materials, including sample demographics, descriptive statistics, and regression coefficients.
In \ref{sec:surveys} we provide details on our survey participants,
in \ref{sec:partisan_supp} we describe the partisan metric we used,
in \ref{sec:descriptives_supp} we discuss descriptive results,
in \ref{sec:regressions} we discuss the multivariate regression results,
and in \ref{sec:irb} we describe our IRB study procedures.
Last, we provide supplementary figures in \ref{sec:figures_supp}, and supplementary tables in \ref{sec:tables_supp}.

\subsection{Opinion Surveys}
\label{sec:surveys}

Tables containing the demographics of our participants are available for both 2018 \autoref{tab:demographics2018} and 2020 \autoref{tab:demographics2020}.
Each table compares the demographics of participants who completed the survey to those who installed the extension, and to estimates from the US Census and the ANES.
In total, 11.5\% of the 3,106 survey participants recruited via YouGov installed the extension and provided at least some data in 2018.
In 2020, we offered participants recruited via PureSpectrum the opportunity to opt-in to installing the extension as part of a much larger survey-focused project (see \url{https://covidstates.org}), and 0.4\% of the 189,711 survey participants installed the extension and provided at least some data.

\subsection{Partisan News Metric} 
\label{sec:partisan_supp}

To quantify participants' exposure and engagement with partisan news, we used the \textit{partisan audience bias} scores developed in prior work~\citep{robertson2018auditinga}.
These scores were made using the domain sharing patterns of a large virtual panel of Twitter users who had been linked to US voter registration records.
More specifically, each domain was scored by comparing the relative proportion of registered Democrats and Republicans who shared it on Twitter.
The resulting scores range from -1 (shared only by Democrats) to 1 (shared only by Republicans). 
See \autoref{tab:partisan_example} for example news domains across the partisanship continuum.

\subsection{Descriptive Results}
\label{sec:descriptives_supp}

In \autoref{tab:counts} we report overall counts for each dataset, including the number of participants, the number of searches conducted, and the number of URLs we observed in each.
In \autoref{tab:search_engine_usage} we examine participants' use of popular search engines and show that Google handled a substantial majority of our participants' web searches in both study waves and across all engagement datasets.
In Tables \ref{tab:user_level} and \ref{tab:news} we compare key descriptive statistics across data types and study waves.

\begin{figure}[!htb]
    \centering
    \includegraphics[width=1\textwidth]{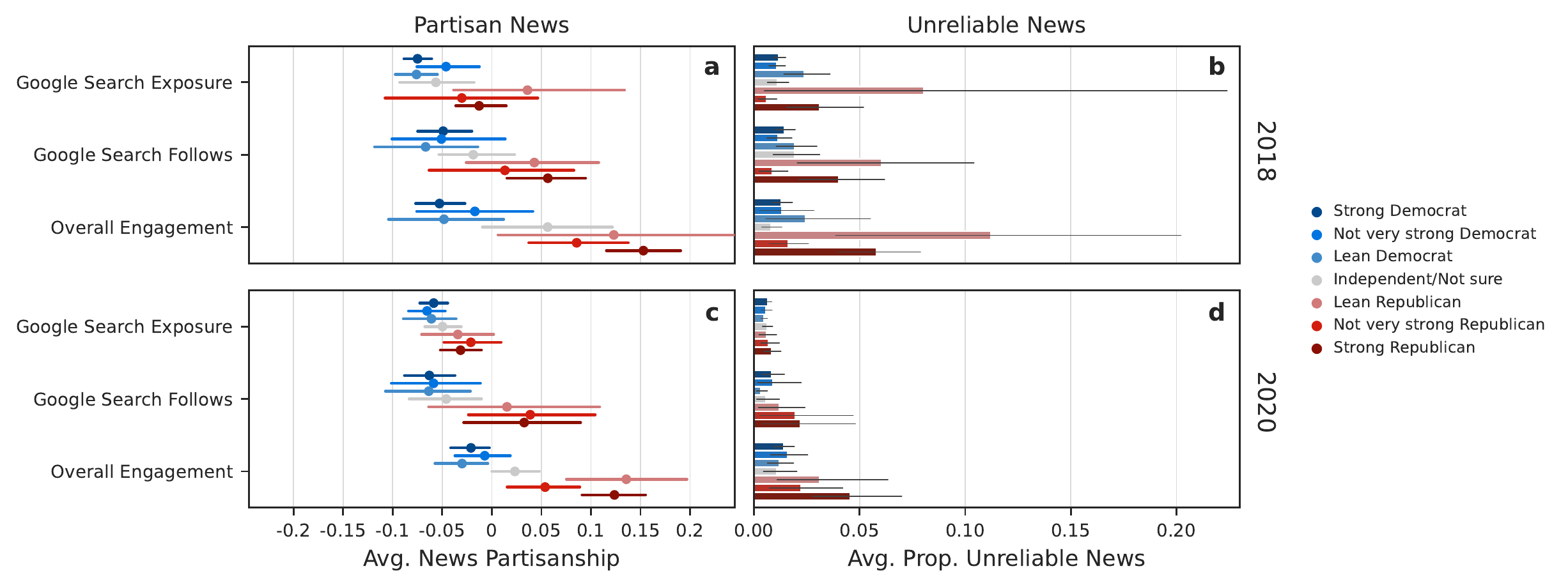}
    \caption{
    Average exposure, follows, and overall engagement (Y-axis) with
    partisan (left X-axis) and unreliable news (right X-axis) by study wave (figure rows) 
    and 7-point PID clustered at the participant-level.
    This figure complements Figure 1 in the main text by displaying all 7-point PID groups instead of the subset that is the focus of this study.
    Error bars indicate 95\% confidence intervals (CI).
    }
    \label{fig:partisan_and_unreliable_news_all_groups}
\end{figure}

For the key descriptive statistics---partisan and unreliable news---we checked for statistically significant differences by political identity and age group. 
To do this, we used the Kruskal-Wallis (KW) test, a non-parametric test of differences among three or more groups. 
We used this nonparametric test because of the heterogeneity we observed in the distribution of user-level averages.
We report the results of all the KW tests we conducted in \autoref{tab:kruskal_tests},
and show differences in partisan and unreliable news for all 7-point partisan identities in \autoref{fig:partisan_and_unreliable_news_all_groups}.
Last, and in support of our finding regarding the concentration of unreliable news (Figure 2), we calculated the Gini coefficient for unreliable news by study wave and dataset and found that it was the lowest within participants' exposure data in both waves (0.771 in 2018, 0.811 in 2020), and greater (more concentrated) in both their follows (0.801 in 2018, 0.939 in 2020) and overall engagement (0.915 in 2018, 0.893 in 2020).

\begin{figure}[!htb]
    \centering
    \includegraphics[width=1\textwidth]{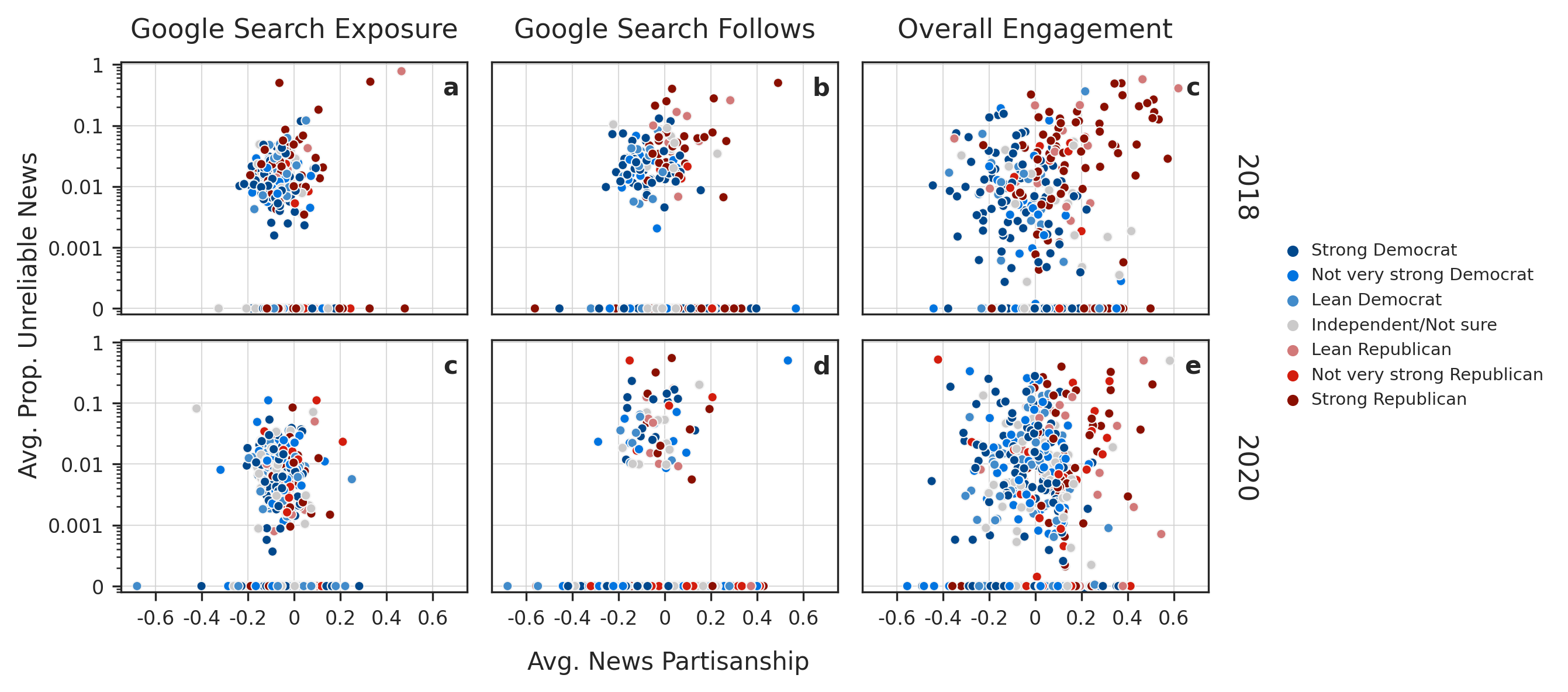}
    \caption{
    The relationship between partisan and unreliable news in our data given participants exposure on Google Search (left), follows from Google Search (middle), and overall engagement (right). 
    This figure complements Figure 3 in the main text by displaying all 7-point PID groups instead of the subset that is the focus of this study.
    }
    \label{fig:partisan_by_unreliable_news_all_groups}
\end{figure}

To evaluate the relationship between partisan and unreliable news consumption, we used Spearman's rank correlation coefficient ($\rho$).
We found no significant correlations between unreliable and partisan news in any dataset, except for exposure in 2020, where we found a small and significant negative correlation ($\rho = -0.164, P < 0.001$).
However, when we examined this correlation within each 7-point partisan ID group (\autoref{tab:spearman_tests}), the correlation for exposure was not significant for any group in either study wave, and significant negative correlations emerged for Strong Democrats while significant positive correlations emerged for Strong Republicans, suggesting a Simpson's paradox.
We show the relationship between partisan and unreliable news for all 7-point partisan identities in \autoref{fig:partisan_by_unreliable_news_all_groups}.

\subsection{Multivariate Regressions}
\label{sec:regressions}

Regression results for each outcome are in:

Partisanship:
\begin{packed_enumerate}
\item Google Search Results: \autoref{tab:bias_search_2018} [2018] and \autoref{tab:bias_search_2020} [2020]
\item Follows from Search: \autoref{tab:bias_follows_2018} [2018] and \autoref{tab:bias_follows_2020} [2020]
\item Browser History: \autoref{tab:bias_browse_2018} [2018] and \autoref{tab:bias_browse_2020} [2020]
\end{packed_enumerate}

Unreliable News:
\begin{packed_enumerate}
\item Google Search Results: \autoref{tab:unreliable_search_2018} [2018] and \autoref{tab:unreliable_search_2020} [2020]
\item Follows from Search: \autoref{tab:unreliable_follows_2018} [2018] and \autoref{tab:unreliable_follows_2020} [2020]
\item Browser History: \autoref{tab:unreliable_browse_2018} [2018] and \autoref{tab:unreliable_browse_2020} [2020]
\end{packed_enumerate}

\subsection{IRB Study Procedures}
\label{sec:irb}

The data we collected from participants is sensitive, thus we took the privacy of subjects and the security of their data contributions seriously. Our study protocol, which was approved by our IRB, included the following (lightly redacted) language about data privacy and security.

\begin{quote}
    All data that is collected from our survey and from participants' browsers will be stripped of personally identifiable information to the best of our ability. Only the research team cited above will have access to the collected data, and it will only be used for the purposes described in this application. Since [survey company] will administer the survey, they will have access to those data, but they will never see or store any of the highly personal data collected through our browser extension. Participants will be identified using an anonymous identifier that is generated by [survey company], or by us, depending on whether the participant is a volunteer. Unfortunately, there is no way to guarantee anonymity for users who allow us to collect their browsing and search history. Collected URLs may contain user’s personal information (e.g., \texttt{www.google.com/search?query=Bob+Smith}), and there is no systematic way to remove this information, since the data does not conform to any set structure. Similarly, the raw HTML of web pages that we collect could contain personally identifying information, such as the users' names or emails in dynamic locations that would be impossible to systematically remove. However, since these data will never be publicly released, the confidentiality of our participants is ensured.
    
    The raw data collected from users will only be stored on a server located within [authors' university], in the [authors' department]. Our software is designed to upload the crawled data directly to our server over an encrypted connection using HTTPS. The server used for this study will be dedicated to this task: it will not be shared with any other unrelated research projects. None of the data that we collect through our extension will be shared with or stored at [survey company], [recruitment partner], or a crowdsourced marketplace. Our server is kept in a physically secure machine room with restricted keycard and PIN access. Remote access to the server is only available to the research group members via encrypted SSH session.

    Aggregated data may be extracted from the server by research group members at different institutions. However, raw data will remain only on our servers. The server will be configured to securely log all accesses---this will allow us to audit which users have access to which data, as well as who has extracted what data.
\end{quote}

Note that, given the sensitivity of our dataset and IRB restrictions, we will not be able to publicly release our dataset.

\section{Supplementary Figures}
\label{sec:figures_supp}

\begin{figure}[!htb]
    \centering
    \includegraphics[width=1\textwidth]{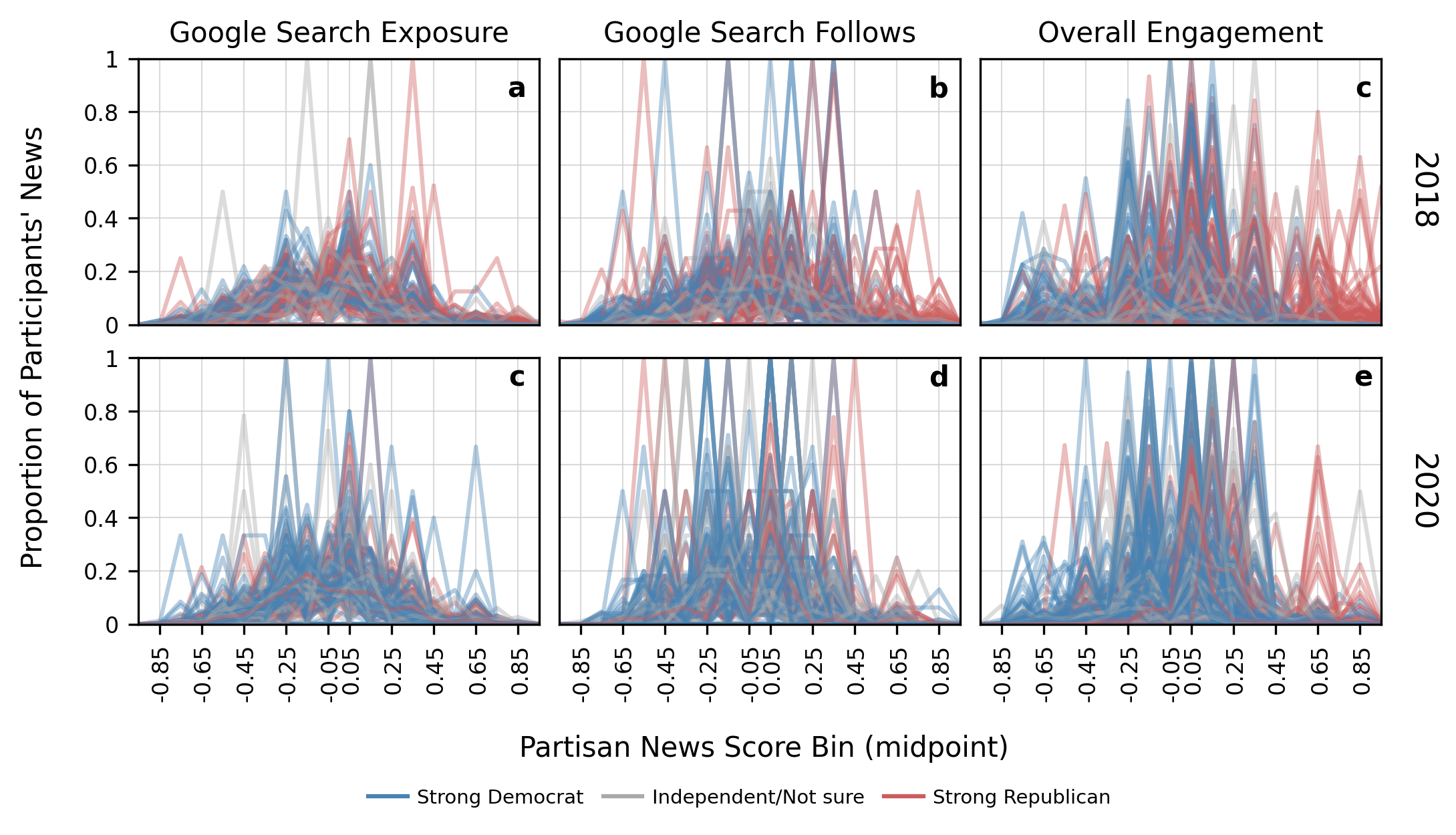}
    \caption{
    Participant-level partisan news distributions for each data type and study wave by 7-point PID. Each line represents the distribution of partisan news sources that a single participant was exposed to in their Google Search results (a, c), followed from those results (b, d), and engaged with overall (c, e). Partisan news scores along the x-axis have been binned in 0.1 point intervals (e.g. -1 to -0.9, -0.9 to -0.8, etc.) with tick labels showing the midpoint of those bins.
    }
    \label{fig:partisan_news_participant}
\end{figure}

\newpage
\FloatBarrier

\section{Supplementary Tables}
\label{sec:tables_supp}

\begin{table}[!htb]
    \begin{adjustwidth}{-0.8cm}{-0.5cm}
    \centering
    \footnotesize
    \resizebox{16cm}{!}{%
    \input{tables/demographics2018}
    }
    \caption{Demographic distributions of our 2018 survey and extension opt-in samples compared to national estimates from the ANES (for 7-Point Party ID) and the US Census.}
    \label{tab:demographics2018}
    \end{adjustwidth}
\end{table}

\begin{table}[!htb]
    \begin{adjustwidth}{-0.8cm}{-0.5cm}
    \centering
    \footnotesize
    \resizebox{16cm}{!}{%
    \input{tables/demographics2020}
    }
    \caption{Demographic distributions of our 2020 survey and extension opt-in samples compared to national estimates from the ANES (for 7-Point Party ID) and the US Census.}
    \label{tab:demographics2020}
    \end{adjustwidth}
\end{table}

\begin{table}[!htb]
    \centering
    \footnotesize
    \input{tables/example_partisanship}
    \caption{Example partisan audience bias scores for news domains from the partisanship measure we used~\cite{robertson2018auditinga}. Included are the total number of shares each score is based on ($N$), as well as the number of shares by Democrats ($N_{dem}$) and Republicans ($N_{rep}$) specifically, in a large virtual Twitter panel.}
    \label{tab:partisan_example}
\end{table}

\begin{table}[!htb]
    \begin{adjustwidth}{-0.8cm}{-0.5cm}
    \centering
    \footnotesize
    \resizebox{18cm}{!}{%
    \input{tables/data_counts}
    }
    \caption{
        Descriptive counts of the number of days we observed participants for, the number of participants, the number of searches they conducted, and the number of URLs, news, and unreliable news they encountered in each of our datasets. Days observed is the number of days between the first and last timestamp across all participants in each dataset.
        For data collected from existing APIs or interfaces, such as Browser History and Google History, we observed a greater number of days because
        they allow collection of data prior to the installation of the extension.
        In contrast, because we built the data collection instruments for Google SERPs and Tab Activity, these can only document participant behavior after the extension has been installed.
    }
    \label{tab:counts}
    \end{adjustwidth}
\end{table}

\begin{table}[!htb]
    \begin{adjustwidth}{-1cm}{-1cm}
    \centering
    \footnotesize
    \input{tables/search_engine_usage}
    \caption{Participants use of popular search engines by study wave and engagement dataset. 
    The total counts for Google are not equal the number of SERPs we collected because the Browser History and Google History engagement datasets include data collected by those services prior to the installation of the browser extension.
    }
    \label{tab:search_engine_usage}
    \end{adjustwidth}
\end{table}

\begin{table}[!htb]
    \begin{adjustwidth}{-0.8cm}{-0.5cm}
    \centering
    \footnotesize
    \resizebox{18cm}{!}{%
    \input{tables/consumption_averages}
    }
    \caption{Participant-level averages and standard deviations for key metric counts across study waves and data types.}
    \label{tab:user_level}
    \end{adjustwidth}
\end{table}

\begin{table}[!htb]
    \begin{adjustwidth}{-0.8cm}{-0.5cm}
    \centering
    \footnotesize
    \resizebox{18cm}{!}{%
    \input{tables/news_consumption}
    }
    \caption{
    News and unreliable news by study wave, data type, and dataset averaged at the participant-level. Here we report the overall percentage of the URLs that came from news and unreliable news domains (using the total URL count as the denominator), as well as the percentage of news that came from unreliable domains (using the total news domain count as the denominator), for the average participant.
    Averages listed under ``Searches'' indicate the percentage of searches that contained at least one news or unreliable news domain, and the ``Unreliable (News)'' column shows the percent of all searches that returned at least one unreliable news domain divided by all searches that returned at least one news domain.
    Our participants' rate of engagement with news was comparable to the rates found in prior work (e.g. 4\% from a study conducted during the 2016 U.S. Presidential election~\cite{peterson2021partisan} and 4.2\% from a study with data spanning 2016 to 2018~\cite{allen2020evaluating}.).
    }
    \label{tab:news}
    \end{adjustwidth}
\end{table}

\begin{table}[!htb]
    \begin{adjustwidth}{-0.8cm}{-0.5cm}
    \centering
    \footnotesize
    \resizebox{18cm}{!}{%
    \input{tables/kruskal_tests}
    }
    \caption{
    Kruskal-Wallis significance tests by study wave, data type, data source, and user grouping. 
    The Partisanship column contains comparisons of partisan news engagement/exposure, and the Unreliable column compares differences in unreliable news engagement/exposure by user groups. 
    Statistical significance level indicated by asterisks---$^{*}P < 0.05$, $^{**}P < 0.01$, $^{***}P < 0.001$---highlighting the absence of statistically significant differences in exposure to unreliable news.
    }
    \label{tab:kruskal_tests}
    \end{adjustwidth}
\end{table}

\begin{table}[!htb]
    \begin{adjustwidth}{-0.8cm}{-0.5cm}
    \centering
    \footnotesize
    \resizebox{18cm}{!}{%
    \input{tables/spearman_tests}
    }
    \caption{
        Spearman rank correlations comparing participants' average news partisanship and proportion of unreliable news by year, data type, data source, and user grouping. 
        Statistical significance levels have been adjusted for multiple hypothesis testing using the Holm method, and levels are indicated by asterisks---$^{*}P < 0.05$, $^{**}P < 0.01$, $^{***}P < 0.001$. 
        Results highlight a Simpson's paradox, and show that engagement with ideologically congruent partisan news is correlated with unreliable news engagement for Strong Democrats and Strong Republicans, depending on the dataset.}
    \label{tab:spearman_tests}
    \end{adjustwidth}
\end{table}

\newpage

\input{regression_tables/bias_search_2018}

\newpage

\input{regression_tables/bias_search_2020}

\newpage

\input{regression_tables/bias_follows_2018}

\newpage

\input{regression_tables/bias_follows_2020}

\newpage

\input{regression_tables/bias_browse_2018}

\newpage

\input{regression_tables/bias_browse_2020}

\newpage

\input{regression_tables/unreliable_search_2018}

\newpage

\input{regression_tables/unreliable_search_2020}

\newpage

\input{regression_tables/unreliable_follows_2018}

\newpage

\input{regression_tables/unreliable_follows_2020}

\newpage

\input{regression_tables/unreliable_browse_2018}

\newpage

\input{regression_tables/unreliable_browse_2020}

%% file: tables/demographics2018.tex
\begin{tabular}{rrrrrrr}
Category & Label  & National & Survey & Opt-in & Survey-National & Opt-in-Survey \\
\midrule
Age Bin & 18 - 24 &    12.6\% &   5.1\% &   6.7\% &           -7.5\% &          1.6\% \\
       & 25 - 44 &    34.2\% &  32.3\% &  33.9\% &           -1.9\% &          1.6\% \\
       & 45 - 64 &    33.9\% &  38.8\% &  38.7\% &            4.9\% &         -0.1\% \\
       & 65+ &    19.3\% &  23.8\% &  20.7\% &            4.5\% &         -3.1\% \\
\cmidrule{2-7}
7-Point Partisan ID & Strong Democrat &    20.8\% &  33.2\% &  36.7\% &           12.3\% &          3.5\% \\
       & Not very strong Democrat &    13.1\% &   6.4\% &   9.5\% &           -6.7\% &          3.1\% \\
       & Lean Democrat &    11.5\% &   5.3\% &   7.3\% &           -6.2\% &          2.0\% \\
       & Independent/Not sure &    14.1\% &  11.5\% &   9.8\% &           -2.6\% &         -1.7\% \\
       & Lean Republican &    11.7\% &   5.6\% &   4.2\% &           -6.1\% &         -1.4\% \\
       & Not very strong Republican &    11.9\% &   5.0\% &   5.0\% &           -6.9\% &          0.1\% \\
       & Strong Republican &    16.9\% &  32.9\% &  27.5\% &           16.1\% &         -5.5\% \\
\cmidrule{2-7}
Gender & Female &    50.8\% &  54.6\% &  44.8\% &            3.9\% &         -9.8\% \\
       & Male &    49.2\% &  45.4\% &  55.2\% &           -3.9\% &          9.8\% \\
\cmidrule{2-7}
Race & Asian &     5.3\% &   2.7\% &   3.9\% &           -2.6\% &          1.2\% \\
       & Black &    12.3\% &   9.8\% &   7.8\% &           -2.5\% &         -1.9\% \\
       & Hispanic &    17.6\% &   8.1\% &   7.3\% &           -9.5\% &         -0.8\% \\
       & Mixed &     2.3\% &   2.4\% &   4.5\% &            0.1\% &          2.1\% \\
       & Native American &     0.7\% &   0.6\% &      - &           -0.1\% &             - \\
       & Other &     0.4\% &   1.4\% &   0.6\% &            1.0\% &         -0.9\% \\
       & White &    61.5\% &  75.0\% &  75.9\% &           13.6\% &          0.9\% \\
\cmidrule{2-7}
Education & No HS &    12.7\% &   4.1\% &   2.8\% &           -8.6\% &         -1.3\% \\
       & High school graduate &    27.3\% &  27.0\% &  16.0\% &           -0.3\% &        -11.0\% \\
       & Some college &    20.8\% &  22.1\% &  20.2\% &            1.3\% &         -2.0\% \\
       & 2-year &     8.3\% &  11.4\% &  11.8\% &            3.1\% &          0.4\% \\
       & 4-year &    19.1\% &  22.8\% &  29.7\% &            3.7\% &          6.9\% \\
       & Post-grad &    11.8\% &  12.7\% &  19.6\% &            0.9\% &          7.0\% \\
\cmidrule{2-7}
Family Income & Less than \$10,000 &     6.7\% &   5.9\% &   4.2\% &           -0.8\% &         -1.7\% \\
      & \$10,000 - \$19,999 &     9.7\% &   8.4\% &  10.6\% &           -1.3\% &          2.3\% \\
      & \$20,000 - \$29,999 &     9.7\% &   9.9\% &  11.8\% &            0.2\% &          1.8\% \\
      & \$30,000 - \$39,999 &     9.3\% &   9.9\% &   9.8\% &            0.6\% &         -0.1\% \\
      & \$40,000 - \$49,999 &     8.5\% &   7.1\% &   7.6\% &           -1.4\% &          0.5\% \\
      & \$50,000 - \$59,999 &     7.7\% &   7.8\% &   7.8\% &            0.1\% &          0.0\% \\
      & \$60,000 - \$99,999 &    22.3\% &  20.3\% &  22.7\% &           -1.9\% &          2.4\% \\
      & \$100,000 - \$149,999 &    14.1\% &  11.1\% &  11.8\% &           -3.0\% &          0.7\% \\
      & \$150,000 - \$199,999 &     5.8\% &   3.3\% &   5.0\% &           -2.5\% &          1.8\% \\
      & \$200,000 or more &     6.3\% &   3.4\% &   3.4\% &           -2.9\% &         -0.0\% \\
      & Prefer not to say &        - &  12.9\% &   5.3\% &               - &         -7.6\% \\
\end{tabular}

%% file: tables/demographics2020.tex
\begin{tabular}{rrrrrrr}
Category & Label  & National & Survey & Opt-in & Survey-National & Opt-in-Survey \\
\midrule
Age Bin & 18 - 24 &    12.6\% &  16.0\% &  10.7\% &            3.4\% &         -5.4\% \\
              & 25 - 44 &    34.2\% &  45.8\% &  37.0\% &           11.5\% &         -8.8\% \\
              & 45 - 64 &    33.9\% &  26.7\% &  34.3\% &           -7.1\% &          7.5\% \\
              & 65+ &    19.3\% &  11.4\% &  18.1\% &           -7.8\% &          6.6\% \\
\cmidrule{2-7}
7-Point Partisan ID & Strong Democrat &    20.8\% &  20.7\% &  27.5\% &           -0.2\% &          6.9\% \\
              & Not very strong Democrat &    13.1\% &  14.7\% &  16.0\% &            1.6\% &          1.3\% \\
              & Lean Democrat &    11.5\% &   9.1\% &  12.8\% &           -2.4\% &          3.7\% \\
              & Independent/Not sure &    14.1\% &  21.1\% &  16.4\% &            7.0\% &         -4.7\% \\
              & Lean Republican &    11.7\% &   6.6\% &   5.4\% &           -5.1\% &         -1.2\% \\
              & Not very strong Republican &    11.9\% &  11.6\% &   9.3\% &           -0.3\% &         -2.3\% \\
              & Strong Republican &    16.9\% &  16.3\% &  12.6\% &           -0.6\% &         -3.7\% \\
\cmidrule{2-7}              
Gender & Female &    50.8\% &  67.3\% &  58.3\% &           16.5\% &         -8.9\% \\
              & Male &    49.2\% &  32.7\% &  41.7\% &          -16.5\% &          8.9\% \\
\cmidrule{2-7}
Race & Asian &     5.3\% &   5.5\% &   5.1\% &            0.3\% &         -0.4\% \\
              & Black or African-American &    12.3\% &  10.5\% &  12.4\% &           -1.8\% &          1.9\% \\
              & Latino &    17.6\% &   7.5\% &   7.3\% &          -10.1\% &         -0.3\% \\
              & Other race not listed &     3.4\% &   4.0\% &   3.4\% &            0.7\% &         -0.6\% \\
              & White &    61.5\% &  72.4\% &  71.8\% &           11.0\% &         -0.6\% \\

\cmidrule{2-7}
Education & Less than a four-year degree &    69.1\% &  59.3\% &  55.0\% &           -9.8\% &         -4.3\% \\
              & Four-year degree or higher &    30.9\% &  40.7\% &  45.0\% &            9.8\% &          4.3\% \\
\cmidrule{2-7}
Family Income & Under \$15,000 &    11.5\% &  12.7\% &   9.3\% &            1.1\% &         -3.4\% \\
              & \$15,000 - \$24,999 &     9.8\% &  11.3\% &  12.9\% &            1.5\% &          1.6\% \\
              & \$25,000 - \$34,999 &     9.5\% &  10.2\% &  12.3\% &            0.6\% &          2.1\% \\
              & \$35,000 - \$49,999 &    13.0\% &  11.8\% &  13.8\% &           -1.2\% &          2.0\% \\
              & \$50,000 - \$74,999 &    17.7\% &  16.2\% &  18.4\% &           -1.4\% &          2.2\% \\
              & \$75,000 - \$99,999 &    12.3\% &  11.3\% &   9.7\% &           -1.1\% &         -1.6\% \\
              & \$100,000 - \$149,999 &    14.1\% &  11.0\% &  12.5\% &           -3.1\% &          1.6\% \\
              & \$150,000 - \$199,999 &     5.8\% &   4.4\% &   4.0\% &           -1.5\% &         -0.3\% \\
              & \$200,000 and over &     6.3\% &   4.3\% &   3.8\% &           -2.0\% &         -0.5\% \\
              & Refused/NA &        - &   7.0\% &   3.3\% &               - &         -3.7\% \\
\end{tabular}

%% file: tables/example_partisanship.tex
\begin{tabular}{lrrrr}
Domain & Score & $N$ & $N_{dem}$ & $N_{rep}$ \\
\toprule
dailykos.com      & -0.705153                 & 12,905                 & 11,960                      & 945                        \\
msnbc.com              & -0.624255                 & 7,071                  & 6,395                       & 676                        \\
 nytimes.com           & -0.260241                 & 149,831                & 118,142                     & 31,689                      \\
huffingtonpost.com    & -0.238265                 & 121,393                & 94,757                      & 26,636                      \\
cnn.com                & -0.118313                 & 98,317                 & 72,278                      & 26,039                      \\
 wsj.com              & 0.010648                  & 64,430                 & 43,926                      & 20,504                      \\
foxnews.com           & 0.607867                  & 38,271                 & 13,318                      & 24,953                      \\
donaldjtrump.com       & 0.693589                  & 2,200                  & 624                        & 1,576                       \\
breitbart.com          & 0.741883                  & 6,783                  & 1,661                       & 5,122                       \\
infowars.com           & 0.781725                  & 3,363                  & 711                        & 2,652                       \\
conservativereview.com & 0.903558                  & 1,563                  & 156                        & 1,407     \\
\bottomrule
\end{tabular}

%% file: tables/data_counts.tex
\begin{tabular}{lllrrllll}
     &      &      & Days     &              &          &      &      &            \\
Year & Type & Data & Observed & Participants & Searches & URLs & News & Unreliable \\
\midrule
2018 & Google Search Exposure & Google SERPs & 73 & 275 & 102,114 & 1,245,155 & 215,699 & 3,932 \\
 & Google Search Follows & Google History & 253 & 262 & - & 279,680 & 22,946 & 661 \\
 & Overall Engagement & Browser History & 253 & 333 & - & 14,677,297 & 994,032 & 29,081 \\
 &  & Google History & 253 & 271 & - & 4,807,758 & 405,596 & 16,754 \\
\cmidrule{2-9}
2020 & Google Search Exposure & Google SERPs & 205 & 459 & 226,035 & 3,654,829 & 586,803 & 5,184 \\
 & Google Search Follows & Tab Activity & 207 & 459 & - & 212,796 & 13,088 & 195 \\
 & Overall Engagement & Browser History & 299 & 688 & - & 31,202,830 & 1,862,011 & 13,209 \\
 &  & Tab Activity & 207 & 597 & - & 20,260,394 & 1,897,933 & 11,880 \\
\bottomrule
\end{tabular}

%% file: tables/search_engine_usage.tex
\begin{tabular}{rrrrrrrr}
&  & \multicolumn{2}{r}{Browser History} & \multicolumn{2}{r}{Google History} & \multicolumn{2}{l}{Tab Activity} \\
\cmidrule(lr){3-4}
\cmidrule(lr){5-6} 
\cmidrule(lr){7-8} 
Year & Search Engine & $N$ & $P$ & $N$ & $P$ & $N$ & $P$ \\
\midrule
2018 & Google & 318,889 & 74.2\% & 518,856 & 91.6\% & - & - \\
& Bing & 93,274 & 21.7\% & 44,273 & 7.8\% & - & - \\
& Yahoo & 9,139 & 2.1\% & 3,064 & 0.5\% & - & - \\
& DuckDuckGo & 8,325 & 1.9\% & 69 & 0.0\% & - & - \\
\cmidrule{2-8}
2020 & Google & 681,359 & 68.6\% & - & - & 306,833 & 64.6\% \\
& Bing & 262,126 & 26.4\% & - & - & 144,859 & 30.5\% \\
& Yahoo & 7,309 & 0.7\% & - & - & 12,137 & 2.6\% \\
& DuckDuckGo & 41,726 & 4.2\% & - & - & 11,262 & 2.4\% \\
\bottomrule
\end{tabular}

%% file: tables/consumption_averages.tex
\begin{tabular}{llrrrrrrrrr}
& &               & Searches &    & URLs  &      & News  &      & Unreliable & \\
\cmidrule(lr){4-5}
\cmidrule(lr){6-7} 
\cmidrule(lr){8-9} 
\cmidrule(lr){10-11}
Year & Type & Data & $\mu$ & $SD$ & $\mu$ & $SD$ & $\mu$ & $SD$ & $\mu$ & $SD$ \\
\midrule
2018 & Google Search Exposure & Google SERPs & 371.3 & 499.7 & 4,527.8 & 6,660.8 & 784.4 & 1,489.5 & 14.3 & 35.3 \\
     & Google Search Follows & Google History & - & - & 1,067.5 & 1,398.3 & 87.6 & 136.0 & 2.5 & 6.9 \\
     & Overall Engagement & Browser History & - & - & 44,076.0 & 49,388.1 & 2,985.1 & 9,625.4 & 87.3 & 467.0 \\
     &  & Google History & - & - & 17,740.8 & 27,620.9 & 1,496.7 & 2,624.3 & 61.8 & 212.2 \\
\cmidrule{2-11}
2020 & Google Search Exposure & Google SERPs & 492.5 & 811.4 & 7,962.6 & 13,594.3 & 1,278.4 & 3,037.2 & 11.3 & 30.1 \\
     & Google Search Follows & Tab Activity & - & - & 463.6 & 791.5 & 28.5 & 100.3 & 0.4 & 2.3 \\
     & Overall Engagement & Browser History & - & - & 45,353.0 & 54,583.6 & 2,706.4 & 14,323.9 & 19.2 & 80.0 \\
     &  & Tab Activity & - & - & 33,937.0 & 49,347.3 & 3,179.1 & 16,022.8 & 19.9 & 78.3 \\
\bottomrule
\end{tabular}

%% file: tables/news_consumption.tex
\begin{tabular}{lllllllll}
     &      &      & \multicolumn{3}{l}{URLs}       & \multicolumn{3}{l}{Searches} \\
\cmidrule(lr){4-6}
\cmidrule(lr){7-9}
     &      &      & News & Unreliable & Unreliable & News & Unreliable & Unreliable \\
Year & Type & Data &      &            & (News)     &      &            & (News)    \\
\midrule
2018 & Google Search Exposure & Google SERPs & 14.31\% & 0.12\% & 2.05\% & 45.61\% & 2.46\% & 4.79\% \\
 & Google Search Follows & Google History & 8.10\% & 0.23\% & 2.36\% & - & - & - \\
 & Overall Engagement & Browser History & 7.11\% & 0.23\% & 3.03\% & - & - & - \\
 &  & Google History & 10.91\% & 0.54\% & 3.64\% & - & - & - \\
\cmidrule{2-9}
2020 & Google Search Exposure & Google SERPs & 14.71\% & 0.04\% & 0.72\% & 52.25\% & 1.18\% & 2.18\% \\
 & Google Search Follows & Tab Activity & 3.47\% & 0.06\% & 2.03\% & - & - & - \\
 & Overall Engagement & Browser History & 3.24\% & 0.04\% & 1.86\% & - & - & - \\
 &  & Tab Activity & 4.36\% & 0.05\% & 0.99\% & - & - & - \\
\end{tabular}

%% file: tables/kruskal_tests.tex
\begin{tabular}{rrrrrllrll}
     &      &      &       & \multicolumn{3}{l}{Partisanship} & \multicolumn{3}{l}{Unreliable} \\
    \cmidrule(lr){5-7} 
    \cmidrule(lr){8-10}
Year & Type & Data & Group & $\chi^2$ & $P$ &   & $\chi^2$ & $P$  & \\
\midrule
2018 & Google Search Exposure & Google SERPs & 7-Point Party ID & 23.25 & 0.00072 & *** & 11.15 & 0.08372 &  \\
 &  &  & Age Bin & 12.64 & 0.00550 & ** & 0.71 & 0.87114 &  \\
 & Google Search Follows & Google History & 7-Point Party ID & 36.27 & 0.00000 & *** & 5.21 & 0.51694 &  \\
 &  &  & Age Bin & 6.80 & 0.07859 &  & 1.85 & 0.60506 &  \\
 & Overall Engagement & Browser History & 7-Point Party ID & 72.22 & 0.00000 & *** & 34.06 & 0.00001 & *** \\
 &  &  & Age Bin & 23.23 & 0.00004 & *** & 19.01 & 0.00027 & *** \\
 &  & Google History & 7-Point Party ID & 66.86 & 0.00000 & *** & 16.26 & 0.01242 & * \\
 &  &  & Age Bin & 11.78 & 0.00819 & ** & 2.45 & 0.48393 &  \\
 \cmidrule{2-9}
2020 & Google Search Exposure & Google SERPs & 7-Point Party ID & 18.97 & 0.00421 & ** & 4.05 & 0.67022 &  \\
 &  &  & Age Bin & 29.48 & 0.00000 & *** & 1.22 & 0.74798 &  \\
 & Google Search Follows & Tab Activity & 7-Point Party ID & 20.70 & 0.00207 & ** & 6.84 & 0.33588 &  \\
 &  &  & Age Bin & 24.67 & 0.00002 & *** & 0.26 & 0.96676 &  \\
 & Overall Engagement & Browser History & 7-Point Party ID & 86.53 & 0.00000 & *** & 18.17 & 0.00581 & ** \\
 &  &  & Age Bin & 29.28 & 0.00000 & *** & 32.01 & 0.00000 & *** \\
 &  & Tab Activity & 7-Point Party ID & 57.72 & 0.00000 & *** & 19.18 & 0.00388 & ** \\
 &  &  & Age Bin & 27.54 & 0.00000 & *** & 34.76 & 0.00000 & *** \\
 \bottomrule
\end{tabular}

%% file: tables/spearman_tests.tex
\begin{tabular}{rrrrrlrrlrrlrrl}
& & & \multicolumn{3}{l}{}        & \multicolumn{3}{l}{Strong}   & \multicolumn{3}{l}{Independent/} & \multicolumn{3}{l}{Strong} \\
& & & \multicolumn{3}{l}{Overall} & \multicolumn{3}{l}{Democrat} & \multicolumn{3}{l}{Not sure}     & \multicolumn{3}{l}{Republican} \\
\cmidrule(lr){4-6}
\cmidrule(lr){7-9}
\cmidrule(lr){10-12}
\cmidrule(lr){13-15}
Year & Type & Data & $\rho$ & $P$ & & $\rho$ & $P$ & & $\rho$ & $P$ & & $\rho$ & $P$ & \\
\midrule
2018 & Google Search Exposure & Google SERPs & -0.035 & 0.567 &  & -0.133 & 1.338 &  & 0.058 & 5.382 &  & -0.030 & 5.599 &  \\
     & Google Search Follows & Google History & -0.062 & 0.331 &  & -0.273 & 0.064 &  & 0.218 & 2.067 &  & 0.029 & 5.770 &  \\
     & Overall Engagement & Browser History & 0.042 & 0.451 &  & -0.380 & 0.000 & *** & -0.093 & 4.377 &  & 0.336 & 0.009 & ** \\
     &  & Google History & 0.112 & 0.067 &  & -0.192 & 0.403 &  & -0.059 & 5.389 &  & 0.377 & 0.009 & ** \\
\cmidrule{2-15}
2020 & Google Search Exposure & Google SERPs & -0.162 & 0.001 & *** & -0.149 & 0.627 &  & -0.140 & 1.648 &  & -0.109 & 3.173 &  \\
     & Google Search Follows & Tab Activity & 0.007 & 0.891 &  & 0.047 & 4.427 &  & 0.002 & 6.914 &  & -0.060 & 5.051 &  \\
     & Overall Engagement & Browser History & -0.018 & 0.648 &  & -0.237 & 0.007 & ** & 0.027 & 5.491 &  & 0.223 & 0.402 &  \\
     &  & Tab Activity & 0.062 & 0.146 &  & -0.140 & 0.577 &  & -0.013 & 6.340 &  & 0.407 & 0.006 & ** \\
\bottomrule
\end{tabular}

%% file: regression_tables/bias_search_2018.tex
\begin{table}
\centering
\scriptsize
\caption{Average Domain Audience Score (Search Results) (2018)}
\resizebox{11cm}{!}{%
\begin{tabular}[t]{lccccc}
\toprule
  & Model 1 & Model 2 & Model 3 & Model 4 & Model 5\\
\midrule
Strong Democrat & -0.025 &  & -0.021 & -0.008 & 0.001\\
 & (0.021) &  & (0.021) & (0.022) & (0.020)\\
Not very strong Democrat & -0.001 &  & 0.005 & 0.013 & 0.027\\
 & (0.027) &  & (0.027) & (0.026) & (0.024)\\
Lean Democrat & -0.024 &  & -0.021 & -0.014 & 0.001\\
 & (0.028) &  & (0.028) & (0.028) & (0.026)\\
Lean Republican & 0.094*** &  & 0.099*** & 0.097*** & 0.093***\\
 & (0.035) &  & (0.035) & (0.035) & (0.032)\\
Not very strong Republican & 0.015 &  & 0.014 & 0.017 & 0.012\\
 & (0.036) &  & (0.036) & (0.036) & (0.033)\\
Strong Republican & 0.029 &  & 0.028 & 0.027 & 0.031\\
 & (0.022) &  & (0.022) & (0.022) & (0.020)\\
Age Group: 25-44 &  & 0.057** & 0.062** & 0.067*** & 0.085***\\
 &  & (0.025) & (0.024) & (0.025) & (0.024)\\
Age Group: 45-64 &  & 0.067*** & 0.066*** & 0.069*** & 0.081***\\
 &  & (0.024) & (0.024) & (0.025) & (0.024)\\
Age Group: 65+ &  & 0.092*** & 0.086*** & 0.094*** & 0.115***\\
 &  & (0.027) & (0.026) & (0.027) & (0.026)\\
High News Interest &  &  &  & -0.019 & -0.016\\
 &  &  &  & (0.014) & (0.014)\\
College &  &  &  & -0.019 & -0.016\\
 &  &  &  & (0.013) & \vphantom{1} (0.012)\\
Female &  &  &  & -0.029** & -0.013\\
 &  &  &  & (0.013) & (0.012)\\
White &  &  &  & 0.003 & 0.007\\
 &  &  &  & (0.014) & (0.013)\\
Query Pivot: 1 &  &  &  &  & -0.251\\
 &  &  &  &  & (0.298)\\
Query Pivot: 2 &  &  &  &  & 0.198\\
 &  &  &  &  & (0.174)\\
Query Pivot: 3 &  &  &  &  & 0.123\\
 &  &  &  &  & (0.233)\\
Query Pivot: 4 &  &  &  &  & -1.011***\\
 &  &  &  &  & (0.180)\\
Query Pivot: 5 &  &  &  &  & -1.326***\\
 &  &  &  &  & (0.255)\\
Query Pivot: 6 &  &  &  &  & -0.647**\\
 &  &  &  &  & (0.327)\\
Query Pivot: 7 &  &  &  &  & -0.054\\
 &  &  &  &  & (0.231)\\
Query Pivot: 8 &  &  &  &  & -0.583**\\
 &  &  &  &  & (0.264)\\
Query Pivot: 9 &  &  &  &  & -0.537*\\
 &  &  &  &  & (0.310)\\
\midrule
Num.Obs. & 270 & 270 & 270 & 270 & 270\\
R2 & 0.085 & 0.045 & 0.122 & 0.152 & 0.335\\
R2 Adj. & 0.064 & 0.034 & 0.092 & 0.109 & 0.276\\
AIC & -476.3 & -470.8 & -481.6 & -483.0 & -530.7\\
BIC & -447.5 & -452.8 & -442.0 & -429.0 & -444.3\\
RMSE & 0.10 & 0.10 & 0.10 & 0.09 & 0.08\\
\bottomrule
\multicolumn{6}{l}{\rule{0pt}{1em}* p $<$ 0.1, ** p $<$ 0.05, *** p $<$ 0.01}\\
\end{tabular}
}
\label{tab:bias_search_2018}
\end{table}

%% file: regression_tables/bias_search_2020.tex
\begin{table}
\scriptsize
\caption{Average Domain Audience Score (Search Results) (2020)}
\centering
\resizebox{11cm}{!}{%
\begin{tabular}[t]{lccccc}
\toprule
  & Model 1 & Model 2 & Model 3 & Model 4 & Model 5\\
\midrule
Strong Democrat & 0.002 &  & -0.002 & 0.007 & 0.004\\
 & (0.012) &  & (0.012) & (0.013) & (0.012)\\
Not very strong Democrat & 0.000 &  & 0.000 & 0.003 & 0.004\\
 & (0.014) &  & (0.014) & (0.014) & (0.013)\\
Lean Democrat & 0.003 &  & 0.000 & 0.002 & 0.002\\
 & (0.014) &  & (0.014) & (0.014) & (0.014)\\
Lean Republican & 0.031 &  & 0.021 & 0.017 & 0.023\\
 & (0.022) &  & (0.022) & (0.022) & (0.021)\\
Not very strong Republican & 0.048*** &  & 0.040** & 0.041** & 0.036**\\
 & (0.017) &  & (0.017) & (0.017) & (0.016)\\
Strong Republican & 0.040** &  & 0.028* & 0.029* & 0.016\\
 & (0.016) &  & (0.016) & (0.016) & (0.015)\\
Age Group: 25-44 &  & 0.021 & 0.018 & 0.018 & 0.013\\
 &  & (0.014) & (0.014) & (0.014) & (0.013)\\
Age Group: 45-64 &  & 0.037*** & 0.029** & 0.026* & 0.018\\
 &  & (0.014) & (0.014) & (0.014) & (0.014)\\
Age Group: 65+ &  & 0.060*** & 0.051*** & 0.048*** & 0.036**\\
 &  & (0.015) & (0.015) & (0.016) & (0.015)\\
High Political Interest &  &  &  & -0.012 & -0.014*\\
 &  &  &  & (0.008) & \vphantom{2} (0.008)\\
College &  &  &  & -0.005 & -0.006\\
 &  &  &  & (0.008) & \vphantom{1} (0.008)\\
Female &  &  &  & -0.014* & -0.012\\
 &  &  &  & (0.008) & (0.008)\\
White &  &  &  & 0.016* & 0.013\\
 &  &  &  & (0.009) & (0.009)\\
Query Pivot: 1 &  &  &  &  & -0.122\\
 &  &  &  &  & \vphantom{1} (0.121)\\
Query Pivot: 2 &  &  &  &  & -0.127\\
 &  &  &  &  & (0.121)\\
Query Pivot: 3 &  &  &  &  & 0.009\\
 &  &  &  &  & (0.126)\\
Query Pivot: 4 &  &  &  &  & -0.036\\
 &  &  &  &  & (0.181)\\
Query Pivot: 5 &  &  &  &  & 0.560**\\
 &  &  &  &  & (0.221)\\
Query Pivot: 6 &  &  &  &  & 0.438*\\
 &  &  &  &  & (0.223)\\
Query Pivot: 7 &  &  &  &  & 0.466***\\
 &  &  &  &  & (0.142)\\
Query Pivot: 8 &  &  &  &  & 0.148\\
 &  &  &  &  & (0.185)\\
Query Pivot: 9 &  &  &  &  & 0.094\\
 &  &  &  &  & (0.155)\\
\midrule
Num.Obs. & 453 & 453 & 453 & 453 & 453\\
R2 & 0.040 & 0.043 & 0.068 & 0.083 & 0.188\\
R2 Adj. & 0.027 & 0.036 & 0.049 & 0.056 & 0.146\\
AIC & -942.1 & -949.1 & -949.1 & -948.6 & -985.5\\
BIC & -909.1 & -928.5 & -903.8 & -886.9 & -886.7\\
RMSE & 0.08 & 0.08 & 0.08 & 0.08 & 0.08\\
\bottomrule
\multicolumn{6}{l}{\rule{0pt}{1em}* p $<$ 0.1, ** p $<$ 0.05, *** p $<$ 0.01}\\
\end{tabular}
}
\label{tab:bias_search_2020}
\end{table}

%% file: regression_tables/bias_follows_2018.tex
\begin{table}

\scriptsize
\caption{Average Domain Audience Score (Follows from Search) (2018)}
\centering
\resizebox{10.75cm}{!}{%
\begin{tabular}[t]{lccccc}
\toprule
  & Model 1 & Model 2 & Model 3 & Model 4 & Model 5\\
\midrule
Strong Democrat & -0.030 &  & -0.030 & -0.010 & -0.006\\
 & (0.032) &  & (0.032) & (0.033) & (0.033)\\
Not very strong Democrat & -0.032 &  & -0.030 & -0.018 & 0.013\\
 & (0.039) &  & (0.040) & (0.040) & (0.039)\\
Lean Democrat & -0.048 &  & -0.044 & -0.029 & -0.024\\
 & (0.043) &  & (0.043) & (0.045) & (0.044)\\
Lean Republican & 0.062 &  & 0.064 & 0.068 & 0.100**\\
 & (0.049) &  & (0.050) & (0.051) & (0.049)\\
Not very strong Republican & 0.032 &  & 0.027 & 0.029 & 0.032\\
 & (0.049) &  & (0.050) & (0.050) & (0.052)\\
Strong Republican & 0.075** &  & 0.071** & 0.073** & 0.085**\\
 & (0.033) &  & (0.033) & (0.034) & (0.034)\\
Age Group: 25-44 &  & 0.057 & 0.048 & 0.056 & 0.069*\\
 &  & (0.038) & (0.037) & (0.038) & \vphantom{1} (0.038)\\
Age Group: 45-64 &  & 0.083** & 0.066* & 0.071* & 0.095**\\
 &  & (0.038) & (0.037) & (0.038) & (0.038)\\
Age Group: 65+ &  & 0.083** & 0.059 & 0.069* & 0.090**\\
 &  & (0.041) & (0.040) & (0.042) & (0.042)\\
High News Interest &  &  &  & -0.031 & -0.036*\\
 &  &  &  & (0.021) & (0.022)\\
College &  &  &  & -0.025 & -0.029\\
 &  &  &  & (0.019) & \vphantom{1} (0.019)\\
Female &  &  &  & -0.031 & -0.031\\
 &  &  &  & (0.019) & (0.019)\\
White &  &  &  & 0.022 & 0.022\\
 &  &  &  & (0.021) & (0.021)\\
Query Pivot: 1 &  &  &  &  & -0.249\\
 &  &  &  &  & (0.326)\\
Query Pivot: 2 &  &  &  &  & 0.295\\
 &  &  &  &  & (0.247)\\
Query Pivot: 3 &  &  &  &  & -0.200\\
 &  &  &  &  & (0.375)\\
Query Pivot: 4 &  &  &  &  & -0.531**\\
 &  &  &  &  & (0.257)\\
Query Pivot: 5 &  &  &  &  & -0.580**\\
 &  &  &  &  & (0.292)\\
Query Pivot: 6 &  &  &  &  & -0.738\\
 &  &  &  &  & (0.455)\\
Query Pivot: 7 &  &  &  &  & -0.214\\
 &  &  &  &  & (0.355)\\
Query Pivot: 8 &  &  &  &  & -0.018\\
 &  &  &  &  & (0.343)\\
Query Pivot: 9 &  &  &  &  & -0.231\\
 &  &  &  &  & (0.462)\\
\midrule
Num.Obs. & 247 & 247 & 247 & 247 & 223\\
R2 & 0.106 & 0.024 & 0.118 & 0.143 & 0.256\\
R2 Adj. & 0.083 & 0.012 & 0.085 & 0.095 & 0.174\\
AIC & -259.0 & -243.4 & -256.6 & -255.6 & -258.7\\
BIC & -230.9 & -225.9 & -218.0 & -203.0 & -176.9\\
RMSE & 0.14 & 0.14 & 0.14 & 0.14 & 0.12\\
\bottomrule
\multicolumn{6}{l}{\rule{0pt}{1em}* p $<$ 0.1, ** p $<$ 0.05, *** p $<$ 0.01}\\
\end{tabular}
}
\label{tab:bias_follows_2018}
\end{table}

%% file: regression_tables/bias_follows_2020.tex
\begin{table}
\scriptsize
\caption{Average Domain Audience Score (Follows from Search) (2020)}
\centering
\resizebox{11cm}{!}{%
\begin{tabular}[t]{lccccc}
\toprule
  & Model 1 & Model 2 & Model 3 & Model 4 & Model 5\\
\midrule
Strong Democrat & -0.020 &  & -0.030 & -0.004 & 0.003\\
 & (0.026) &  & (0.026) & (0.027) & (0.027)\\
Not very strong Democrat & -0.012 &  & -0.019 & -0.013 & 0.001\\
 & (0.030) &  & (0.030) & (0.029) & (0.030)\\
Lean Democrat & -0.017 &  & -0.024 & -0.021 & -0.014\\
 & (0.030) &  & (0.030) & (0.030) & (0.030)\\
Lean Republican & 0.063 &  & 0.034 & 0.025 & 0.046\\
 & (0.047) &  & (0.047) & (0.047) & (0.048)\\
Not very strong Republican & 0.087** &  & 0.069* & 0.062* & 0.070*\\
 & (0.037) &  & (0.036) & (0.036) & (0.036)\\
Strong Republican & 0.080** &  & 0.048 & 0.052 & 0.063*\\
 & (0.033) &  & (0.034) & (0.034) & (0.035)\\
Age Group: 25-44 &  & 0.034 & 0.025 & 0.030 & 0.025\\
 &  & (0.028) & (0.028) & (0.029) & (0.029)\\
Age Group: 45-64 &  & 0.071** & 0.053* & 0.049* & 0.050*\\
 &  & (0.029) & (0.029) & (0.029) & (0.030)\\
Age Group: 65+ &  & 0.130*** & 0.111*** & 0.108*** & 0.105***\\
 &  & (0.031) & (0.031) & (0.031) & (0.033)\\
High Political Interest &  &  &  & -0.050*** & -0.052***\\
 &  &  &  & (0.018) & (0.018)\\
College &  &  &  & -0.018 & -0.018\\
 &  &  &  & (0.017) & (0.018)\\
Female &  &  &  & -0.036** & -0.033*\\
 &  &  &  & (0.018) & (0.019)\\
White &  &  &  & 0.046** & 0.039**\\
 &  &  &  & (0.019) & (0.020)\\
Query Pivot: 1 &  &  &  &  & -0.289\\
 &  &  &  &  & (0.305)\\
Query Pivot: 2 &  &  &  &  & -0.545*\\
 &  &  &  &  & (0.306)\\
Query Pivot: 3 &  &  &  &  & 0.025\\
 &  &  &  &  & (0.318)\\
Query Pivot: 4 &  &  &  &  & 0.757*\\
 &  &  &  &  & (0.450)\\
Query Pivot: 5 &  &  &  &  & -0.524\\
 &  &  &  &  & (0.565)\\
Query Pivot: 6 &  &  &  &  & 0.538\\
 &  &  &  &  & (0.571)\\
Query Pivot: 7 &  &  &  &  & 0.475\\
 &  &  &  &  & (0.429)\\
Query Pivot: 8 &  &  &  &  & -0.001\\
 &  &  &  &  & (0.477)\\
Query Pivot: 9 &  &  &  &  & 0.402\\
 &  &  &  &  & (0.384)\\
\midrule
Num.Obs. & 350 & 350 & 350 & 350 & 350\\
R2 & 0.060 & 0.065 & 0.105 & 0.146 & 0.165\\
R2 Adj. & 0.044 & 0.057 & 0.081 & 0.113 & 0.108\\
AIC & -284.4 & -292.3 & -295.5 & -304.0 & -293.7\\
BIC & -253.5 & -273.0 & -253.1 & -246.1 & -201.1\\
RMSE & 0.16 & 0.16 & 0.15 & 0.15 & 0.15\\
\bottomrule
\multicolumn{6}{l}{\rule{0pt}{1em}* p $<$ 0.1, ** p $<$ 0.05, *** p $<$ 0.01}\\
\end{tabular}
}
\label{tab:bias_follows_2020}
\end{table}

%% file: regression_tables/bias_browse_2018.tex
\begin{table}

\scriptsize
\caption{Average Domain Audience Score (Browser History) (2018)}
\centering
\begin{tabular}[t]{lcccc}
\toprule
  & Model 1 & Model 2 & Model 3 & Model 4\\
\midrule
Strong Democrat & -0.109*** &  & -0.106*** & -0.097***\\
 & (0.034) &  & (0.033) & (0.034)\\
Not very strong Democrat & -0.073* &  & -0.067 & -0.065\\
 & (0.043) &  & (0.042) & (0.043)\\
Lean Democrat & -0.105** &  & -0.107** & -0.100**\\
 & (0.045) &  & (0.045) & (0.046)\\
Lean Republican & 0.067 &  & 0.071 & 0.081\\
 & (0.052) &  & (0.052) & (0.052)\\
Not very strong Republican & 0.029 &  & 0.021 & 0.035\\
 & (0.050) &  & (0.049) & (0.050)\\
Strong Republican & 0.097*** &  & 0.090*** & 0.099***\\
 & (0.035) &  & (0.034) & (0.035)\\
Age Group: 25-44 &  & 0.102** & 0.092** & 0.085**\\
 &  & (0.042) & (0.038) & (0.038)\\
Age Group: 45-64 &  & 0.145*** & 0.114*** & 0.111***\\
 &  & (0.041) & (0.038) & (0.039)\\
Age Group: 65+ &  & 0.179*** & 0.151*** & 0.150***\\
 &  & (0.044) & (0.040) & (0.041)\\
High News Interest &  &  &  & -0.017\\
 &  &  &  & (0.021)\\
College &  &  &  & 0.021\\
 &  &  &  & \vphantom{1} (0.019)\\
Female &  &  &  & -0.035*\\
 &  &  &  & (0.019)\\
White &  &  &  & -0.026\\
 &  &  &  & (0.022)\\
\midrule
Num.Obs. & 319 & 319 & 319 & 319\\
R2 & 0.229 & 0.059 & 0.266 & 0.281\\
R2 Adj. & 0.214 & 0.050 & 0.245 & 0.250\\
AIC & -232.7 & -175.0 & -242.4 & -240.9\\
BIC & -202.6 & -156.2 & -200.9 & -184.4\\
RMSE & 0.16 & 0.18 & 0.16 & 0.16\\
\bottomrule
\multicolumn{5}{l}{\rule{0pt}{1em}* p $<$ 0.1, ** p $<$ 0.05, *** p $<$ 0.01}\\
\end{tabular}
\normalsize

\label{tab:bias_browse_2018}

\end{table}

%% file: regression_tables/bias_browse_2020.tex
\begin{table}

\scriptsize
\caption{Average Domain Audience Score (Browser History) (2020)}
\centering
\begin{tabular}[t]{lcccc}
\toprule
  & Model 1 & Model 2 & Model 3 & Model 4\\
\midrule
Strong Democrat & -0.042** &  & -0.047*** & -0.031*\\
 & (0.017) &  & (0.017) & (0.017)\\
Not very strong Democrat & -0.035* &  & -0.037* & -0.032*\\
 & (0.019) &  & (0.019) & (0.019)\\
Lean Democrat & -0.054*** &  & -0.060*** & -0.054***\\
 & (0.020) &  & (0.020) & (0.020)\\
Lean Republican & 0.114*** &  & 0.098*** & 0.094***\\
 & (0.027) &  & (0.027) & (0.027)\\
Not very strong Republican & 0.036 &  & 0.024 & 0.026\\
 & (0.023) &  & (0.023) & (0.022)\\
Strong Republican & 0.093*** &  & 0.075*** & 0.080***\\
 & (0.021) &  & (0.021) & (0.021)\\
Age Group: 25-44 &  & 0.031 & 0.021 & 0.021\\
 &  & (0.020) & (0.019) & (0.019)\\
Age Group: 45-64 &  & 0.082*** & 0.054*** & 0.053***\\
 &  & (0.020) & (0.019) & (0.020)\\
Age Group: 65+ &  & 0.090*** & 0.067*** & 0.068***\\
 &  & (0.022) & (0.021) & (0.021)\\
High Political Interest &  &  &  & -0.028**\\
 &  &  &  & \vphantom{2} (0.011)\\
College &  &  &  & -0.015\\
 &  &  &  & \vphantom{1} (0.011)\\
Female &  &  &  & -0.040***\\
 &  &  &  & (0.011)\\
White &  &  &  & 0.021\\
 &  &  &  & (0.013)\\
\midrule
Num.Obs. & 655 & 655 & 655 & 655\\
R2 & 0.130 & 0.047 & 0.152 & 0.177\\
R2 Adj. & 0.122 & 0.042 & 0.140 & 0.161\\
AIC & -722.8 & -668.7 & -733.1 & -745.2\\
BIC & -686.9 & -646.3 & -683.7 & -677.9\\
RMSE & 0.14 & 0.14 & 0.14 & 0.13\\
\bottomrule
\multicolumn{5}{l}{\rule{0pt}{1em}* p $<$ 0.1, ** p $<$ 0.05, *** p $<$ 0.01}\\
\end{tabular}
\normalsize

\label{tab:bias_browse_2020}

\end{table}

%% file: regression_tables/unreliable_search_2018.tex
\begin{table}
\scriptsize
\caption{Unreliable URLs (Search Results) (2018)}
\centering
\resizebox{11cm}{!}{%
\begin{tabular}[t]{lccccc}
\toprule
  & Model 1 & Model 2 & Model 3 & Model 4 & Model 5\\
\midrule
Strong Democrat & -0.287 &  & -0.208 & -0.391 & -0.337\\
 & (0.234) &  & (0.233) & (0.247) & (0.223)\\
Not very strong Democrat & -0.239 &  & -0.141 & -0.229 & -0.265\\
 & (0.294) &  & (0.293) & (0.296) & (0.265)\\
Lean Democrat & 0.307 &  & 0.240 & 0.025 & 0.039\\
 & (0.289) &  & (0.289) & (0.298) & (0.272)\\
Lean Republican & 0.473 &  & 0.361 & 0.216 & -0.103\\
 & (0.387) &  & (0.391) & (0.392) & (0.376)\\
Not very strong Republican & -0.581 &  & -0.554 & -0.732* & -0.504\\
 & (0.424) &  & (0.424) & (0.427) & (0.385)\\
Strong Republican & 0.482** &  & 0.435* & 0.192 & 0.124\\
 & (0.241) &  & (0.240) & (0.246) & (0.222)\\
Age Group: 25-44 &  & 0.315 & 0.273 & 0.128 & 0.169\\
 &  & (0.242) & (0.239) & (0.245) & (0.230)\\
Age Group: 45-64 &  & 0.296 & 0.227 & -0.024 & 0.060\\
 &  & (0.240) & (0.238) & (0.245) & (0.234)\\
Age Group: 65+ &  & 0.975*** & 0.700*** & 0.370 & 0.386\\
 &  & (0.263) & (0.261) & (0.275) & (0.260)\\
High News Interest &  &  &  & 0.410*** & 0.304**\\
 &  &  &  & (0.158) & (0.147)\\
College &  &  &  & -0.090 & -0.059\\
 &  &  &  & (0.135) & (0.125)\\
Female &  &  &  & -0.160 & -0.046\\
 &  &  &  & (0.133) & (0.128)\\
White &  &  &  & 0.168 & 0.346**\\
 &  &  &  & (0.155) & (0.146)\\
Query Pivot: 1 &  &  &  &  & -1.830\\
 &  &  &  &  & (3.770)\\
Query Pivot: 2 &  &  &  &  & -1.253\\
 &  &  &  &  & (1.924)\\
Query Pivot: 3 &  &  &  &  & 0.701\\
 &  &  &  &  & (2.845)\\
Query Pivot: 4 &  &  &  &  & 4.804**\\
 &  &  &  &  & (2.152)\\
Query Pivot: 5 &  &  &  &  & -19.642***\\
 &  &  &  &  & (3.372)\\
Query Pivot: 6 &  &  &  &  & 7.715*\\
 &  &  &  &  & (4.078)\\
Query Pivot: 7 &  &  &  &  & 4.085\\
 &  &  &  &  & (2.754)\\
Query Pivot: 8 &  &  &  &  & -8.800***\\
 &  &  &  &  & (3.281)\\
Query Pivot: 9 &  &  &  &  & 6.523*\\
 &  &  &  &  & (3.905)\\
\midrule
AIC & 1468.3 & 1474.2 & 1465.4 & 1461.2 & 1416.7\\
Num.Obs & 275 & 275 & 275 & 275 & 273\\
Dev & 293.996 & 299.146 & 288.707 & 279.382 & 264.118\\
Null.Dev & 325.698 & 318.598 & 329.523 & 332.583 & 388.665\\
\bottomrule
\multicolumn{6}{l}{\rule{0pt}{1em}* p $<$ 0.1, ** p $<$ 0.05, *** p $<$ 0.01}\\
\end{tabular}
}
\label{tab:unreliable_search_2018}
\end{table}

%% file: regression_tables/unreliable_search_2020.tex
\begin{table}
\centering
\scriptsize
\caption{Unreliable URLs (Search Results) (2020)}
\resizebox{11cm}{!}{%
\begin{tabular}[t]{lccccc}
\toprule
  & Model 1 & Model 2 & Model 3 & Model 4 & Model 5\\
\midrule
Strong Democrat & 0.019 &  & 0.061 & 0.003 & 0.018\\
 & (0.179) &  & (0.181) & (0.192) & (0.191)\\
Not very strong Democrat & 0.078 &  & 0.123 & 0.100 & 0.092\\
 & (0.205) &  & (0.208) & (0.210) & (0.210)\\
Lean Democrat & -0.006 &  & -0.012 & 0.010 & 0.003\\
 & (0.206) &  & (0.206) & (0.208) & (0.209)\\
Lean Republican & -0.197 &  & -0.204 & -0.188 & -0.209\\
 & (0.339) &  & (0.343) & (0.348) & (0.350)\\
Not very strong Republican & 0.015 &  & 0.043 & 0.052 & 0.152\\
 & (0.252) &  & (0.256) & (0.258) & (0.259)\\
Strong Republican & -0.128 &  & -0.081 & -0.098 & 0.071\\
 & (0.229) &  & (0.237) & (0.243) & (0.247)\\
Age Group: 25-44 &  & 0.236 & 0.273 & 0.270 & 0.292\\
 &  & (0.193) & (0.197) & (0.202) & (0.204)\\
Age Group: 45-64 &  & 0.108 & 0.141 & 0.157 & 0.230\\
 &  & (0.196) & (0.201) & (0.209) & (0.213)\\
Age Group: 65+ &  & 0.044 & 0.077 & 0.104 & 0.139\\
 &  & (0.215) & (0.221) & (0.228) & (0.236)\\
High Political Interest &  &  &  & 0.173 & 0.212\\
 &  &  &  & (0.129) & (0.129)\\
College &  &  &  & -0.005 & 0.000\\
 &  &  &  & (0.121) & (0.123)\\
Female &  &  &  & 0.075 & 0.026\\
 &  &  &  & (0.126) & (0.133)\\
White &  &  &  & -0.135 & -0.168\\
 &  &  &  & (0.137) & (0.139)\\
Query Pivot: 1 &  &  &  &  & 1.147\\
 &  &  &  &  & (2.195)\\
Query Pivot: 2 &  &  &  &  & -2.479\\
 &  &  &  &  & (2.147)\\
Query Pivot: 3 &  &  &  &  & 0.258\\
 &  &  &  &  & (2.372)\\
Query Pivot: 4 &  &  &  &  & 2.765\\
 &  &  &  &  & (3.348)\\
Query Pivot: 5 &  &  &  &  & -8.495**\\
 &  &  &  &  & (4.328)\\
Query Pivot: 6 &  &  &  &  & -3.309\\
 &  &  &  &  & (4.240)\\
Query Pivot: 7 &  &  &  &  & 2.410\\
 &  &  &  &  & (2.779)\\
Query Pivot: 8 &  &  &  &  & 1.503\\
 &  &  &  &  & (3.458)\\
Query Pivot: 9 &  &  &  &  & -2.801\\
 &  &  &  &  & (2.862)\\
\midrule
AIC & 2140.5 & 2133.2 & 2143.8 & 2149.1 & 2151.9\\
Num.Obs & 459 & 459 & 459 & 459 & 459\\
Dev & 451.807 & 455.675 & 449.234 & 444.891 & 435.827\\
Null.Dev & 453.043 & 458.109 & 453.161 & 451.574 & 457.648\\
\bottomrule
\multicolumn{6}{l}{\rule{0pt}{1em}* p $<$ 0.1, ** p $<$ 0.05, *** p $<$ 0.01}\\
\end{tabular}
}
\label{tab:unreliable_search_2020}
\end{table}

%% file: regression_tables/unreliable_follows_2018.tex
\begin{table}
\scriptsize
\centering
\caption{Unreliable URLs (Follows from Search) (2018)}
\resizebox{11cm}{!}{%
\begin{tabular}[t]{lccccc}
\toprule
  & Model 1 & Model 2 & Model 3 & Model 4 & Model 5\\
\midrule
Strong Democrat & 2.491*** &  & 2.277*** & -0.173 & 0.046\\
 & (0.671) &  & (0.660) & (0.651) & (0.634)\\
Not very strong Democrat & -0.529 &  & -0.594 & -1.648** & -0.965\\
 & (0.834) &  & (0.824) & (0.791) & (0.749)\\
Lean Democrat & 0.420 &  & 0.940 & -0.625 & 0.446\\
 & (0.865) &  & (0.860) & (0.852) & (0.810)\\
Lean Republican & -1.194 &  & -0.480 & -1.189 & -0.215\\
 & (0.995) &  & (1.001) & (0.960) & (0.940)\\
Not very strong Republican & -1.068 &  & -0.717 & -2.206** & -0.963\\
 & (1.053) &  & (1.027) & (1.017) & (0.969)\\
Strong Republican & -0.082 &  & 0.102 & -0.890 & -0.601\\
 & (0.701) &  & (0.691) & (0.667) & (0.651)\\
Age Group: 25-44 &  & 1.696** & 1.628** & 1.701** & 0.203\\
 &  & (0.743) & (0.733) & (0.728) & (0.719)\\
Age Group: 45-64 &  & 2.832*** & 1.528** & 0.844 & -0.678\\
 &  & (0.736) & (0.726) & (0.729) & (0.719)\\
Age Group: 65+ &  & -0.832 & -0.474 & -0.525 & -2.079***\\
 &  & (0.807) & (0.788) & (0.800) & (0.792)\\
High News Interest &  &  &  & 1.232*** & 0.550\\
 &  &  &  & (0.404) & (0.410)\\
College &  &  &  & 0.575 & 0.648*\\
 &  &  &  & (0.363) & (0.358)\\
Female &  &  &  & 1.934*** & 1.373***\\
 &  &  &  & (0.361) & (0.365)\\
White &  &  &  & 0.557 & 1.047**\\
 &  &  &  & (0.411) & (0.410)\\
Query Pivot: 1 &  &  &  &  & 31.182***\\
 &  &  &  &  & (5.429)\\
Query Pivot: 2 &  &  &  &  & -2.237\\
 &  &  &  &  & (4.561)\\
Query Pivot: 3 &  &  &  &  & -12.360*\\
 &  &  &  &  & (6.951)\\
Query Pivot: 4 &  &  &  &  & 10.141*\\
 &  &  &  &  & (5.359)\\
Query Pivot: 5 &  &  &  &  & 1.297\\
 &  &  &  &  & (5.703)\\
Query Pivot: 6 &  &  &  &  & 10.845\\
 &  &  &  &  & (8.784)\\
Query Pivot: 7 &  &  &  &  & 6.097\\
 &  &  &  &  & (7.256)\\
Query Pivot: 8 &  &  &  &  & -26.389***\\
 &  &  &  &  & (6.899)\\
Query Pivot: 9 &  &  &  &  & -25.877***\\
 &  &  &  &  & (9.347)\\
\midrule
AIC & 1112.0 & 1113.6 & 1103.7 & 1089.2 & 1018.1\\
Num.Obs & 259 & 259 & 259 & 259 & 231\\
Dev & 235.749 & 235.499 & 229.572 & 224.218 & 195.579\\
Null.Dev & 278.822 & 270.268 & 287.954 & 308.394 & 328.509\\
\bottomrule
\multicolumn{6}{l}{\rule{0pt}{1em}* p $<$ 0.1, ** p $<$ 0.05, *** p $<$ 0.01}\\
\end{tabular}
}
\label{tab:unreliable_follows_2018}
\end{table}

%% file: regression_tables/unreliable_follows_2020.tex
\begin{table}
\centering
\scriptsize
\caption{Unreliable URLs (Follows from Search) (2020)}
\resizebox{11cm}{!}{%
\begin{tabular}[t]{lccccc}
\toprule
  & Model 1 & Model 2 & Model 3 & Model 4 & Model 5\\
\midrule
Strong Democrat & 0.235 &  & 0.090 & 0.084 & 0.424\\
 & (0.479) &  & (0.485) & (0.512) & (0.532)\\
Not very strong Democrat & -0.180 &  & -0.191 & -0.242 & -0.064\\
 & (0.570) &  & (0.564) & (0.577) & (0.587)\\
Lean Democrat & -0.091 &  & -0.296 & -0.250 & -0.256\\
 & (0.552) &  & (0.558) & (0.572) & (0.600)\\
Lean Republican & 0.556 &  & 0.459 & 0.455 & 0.927\\
 & (0.781) &  & (0.783) & (0.806) & (0.826)\\
Not very strong Republican & 0.967 &  & 0.607 & 0.552 & 1.057\\
 & (0.621) &  & (0.634) & (0.655) & (0.670)\\
Strong Republican & 1.402*** &  & 1.012* & 1.008* & 1.136*\\
 & (0.538) &  & (0.557) & (0.582) & (0.641)\\
Age Group: 25-44 &  & 0.349 & 0.314 & 0.326 & 0.453\\
 &  & (0.589) & (0.592) & (0.613) & (0.632)\\
Age Group: 45-64 &  & 1.342** & 0.943 & 1.026* & 1.114*\\
 &  & (0.575) & (0.580) & (0.602) & (0.641)\\
Age Group: 65+ &  & 0.869 & 0.682 & 0.752 & 0.825\\
 &  & (0.627) & (0.625) & (0.644) & (0.683)\\
High Political Interest &  &  &  & 0.170 & 0.181\\
 &  &  &  & (0.337) & (0.354)\\
College &  &  &  & -0.224 & -0.324\\
 &  &  &  & (0.316) & (0.339)\\
Female &  &  &  & 0.075 & -0.059\\
 &  &  &  & (0.327) & (0.368)\\
White &  &  &  & -0.215 & -0.315\\
 &  &  &  & (0.366) & (0.385)\\
Query Pivot: 1 &  &  &  &  & -1.207\\
 &  &  &  &  & (5.994)\\
Query Pivot: 2 &  &  &  &  & 8.407\\
 &  &  &  &  & (6.174)\\
Query Pivot: 3 &  &  &  &  & -16.844**\\
 &  &  &  &  & (6.910)\\
Query Pivot: 4 &  &  &  &  & 15.876*\\
 &  &  &  &  & (9.173)\\
Query Pivot: 5 &  &  &  &  & -9.048\\
 &  &  &  &  & (11.923)\\
Query Pivot: 6 &  &  &  &  & 12.114\\
 &  &  &  &  & (12.096)\\
Query Pivot: 7 &  &  &  &  & -6.306\\
 &  &  &  &  & (8.250)\\
Query Pivot: 8 &  &  &  &  & -15.597\\
 &  &  &  &  & (9.737)\\
Query Pivot: 9 &  &  &  &  & 7.532\\
 &  &  &  &  & (8.686)\\
\midrule
AIC & 466.4 & 465.0 & 468.3 & 475.7 & 479.2\\
Num.Obs & 448 & 448 & 448 & 448 & 448\\
Dev & 155.035 & 153.639 & 150.934 & 146.628 & 137.234\\
Null.Dev & 171.611 & 165.052 & 171.538 & 167.513 & 173.145\\
\bottomrule
\multicolumn{6}{l}{\rule{0pt}{1em}* p $<$ 0.1, ** p $<$ 0.05, *** p $<$ 0.01}\\
\end{tabular}
}
\label{tab:unreliable_follows_2020}
\end{table}

%% file: regression_tables/unreliable_browse_2018.tex
\begin{table}

\scriptsize
\caption{Unreliable URLs (Browser History) (2018)}
\centering
\begin{tabular}[t]{lcccc}
\toprule
  & Model 1 & Model 2 & Model 3 & Model 4\\
\midrule
Strong Democrat & 0.265 &  & 0.272 & -0.060\\
 & (0.384) &  & (0.382) & (0.393)\\
Not very strong Democrat & 0.285 &  & 0.242 & -0.067\\
 & (0.471) &  & (0.469) & (0.470)\\
Lean Democrat & 0.727 &  & 1.033** & 0.315\\
 & (0.494) &  & (0.492) & (0.502)\\
Lean Republican & 2.315*** &  & 2.245*** & 1.894***\\
 & (0.553) &  & (0.550) & (0.547)\\
Not very strong Republican & 0.435 &  & 0.483 & 0.372\\
 & (0.561) &  & (0.554) & (0.552)\\
Strong Republican & 1.699*** &  & 1.571*** & 1.268***\\
 & (0.391) &  & (0.389) & (0.392)\\
Age Group: 25-44 &  & -0.237 & -0.041 & -0.226\\
 &  & (0.423) & (0.408) & (0.414)\\
Age Group: 45-64 &  & 0.688* & 0.732* & 0.354\\
 &  & (0.415) & (0.401) & (0.411)\\
Age Group: 65+ &  & 0.991** & 0.781* & 0.470\\
 &  & (0.442) & (0.423) & (0.437)\\
High News Interest &  &  &  & 0.960***\\
 &  &  &  & (0.227)\\
College &  &  &  & -0.200\\
 &  &  &  & (0.204)\\
Female &  &  &  & 0.079\\
 &  &  &  & (0.207)\\
White &  &  &  & -0.074\\
 &  &  &  & (0.235)\\
\midrule
AIC & 2312.5 & 2343.2 & 2304.8 & 2297.1\\
Num.Obs & 333 & 333 & 333 & 333\\
Dev & 335.398 & 341.422 & 331.026 & 324.940\\
Null.Dev & 398.488 & 363.995 & 409.696 & 421.773\\
\bottomrule
\multicolumn{5}{l}{\rule{0pt}{1em}* p $<$ 0.1, ** p $<$ 0.05, *** p $<$ 0.01}\\
\end{tabular}
\normalsize

\label{tab:unreliable_browse_2018}

\end{table}

%% file: regression_tables/unreliable_browse_2020.tex
\begin{table}

\scriptsize
\caption{Unreliable URLs (Browser History) (2020)}
\centering
\begin{tabular}[t]{lcccc}
\toprule
  & Model 1 & Model 2 & Model 3 & Model 4\\
\midrule
Strong Democrat & 0.128 &  & 0.164 & 0.232\\
 & (0.254) &  & (0.256) & (0.262)\\
Not very strong Democrat & 0.212 &  & 0.210 & 0.255\\
 & (0.290) &  & (0.292) & (0.291)\\
Lean Democrat & -0.006 &  & -0.005 & 0.160\\
 & (0.301) &  & (0.301) & (0.299)\\
Lean Republican & 0.656 &  & 0.512 & 0.644\\
 & (0.401) &  & (0.401) & (0.397)\\
Not very strong Republican & 0.668** &  & 0.438 & 0.608*\\
 & (0.336) &  & (0.339) & (0.336)\\
Strong Republican & 1.198*** &  & 1.015*** & 1.141***\\
 & (0.308) &  & (0.312) & (0.311)\\
Age Group: 25-44 &  & 0.491 & 0.437 & 0.523*\\
 &  & (0.309) & (0.309) & (0.311)\\
Age Group: 45-64 &  & 1.316*** & 1.106*** & 1.014***\\
 &  & (0.303) & (0.307) & (0.314)\\
Age Group: 65+ &  & 1.306*** & 1.055*** & 1.067***\\
 &  & (0.325) & (0.327) & (0.333)\\
High Political Interest &  &  &  & 0.441***\\
 &  &  &  & (0.170)\\
College &  &  &  & -0.643***\\
 &  &  &  & (0.161)\\
Female &  &  &  & 0.235\\
 &  &  &  & (0.165)\\
White &  &  &  & -0.080\\
 &  &  &  & (0.191)\\
\midrule
AIC & 3398.4 & 3389.5 & 3384.6 & 3369.6\\
Num.Obs & 688 & 688 & 688 & 688\\
Dev & 599.865 & 603.823 & 597.147 & 592.852\\
Null.Dev & 628.340 & 635.542 & 646.459 & 667.012\\
\bottomrule
\multicolumn{5}{l}{\rule{0pt}{1em}* p $<$ 0.1, ** p $<$ 0.05, *** p $<$ 0.01}\\
\end{tabular}
\normalsize
\label{tab:unreliable_browse_2020}

\end{table}

%% file: refs.bib
@article{allcott2017social,
  title = {Social Media and Fake News in the 2016 Election},
  author = {Allcott, Hunt and Gentzkow, Matthew},
  year = {2017},
  journal = {Journal of Economic Perspectives},
  volume = {31},
  number = {2},
  pages = {211--236},
  issn = {0895-3309},
  doi = {10.1257/jep.31.2.211}
}

@article{allen2020evaluating,
  title = {Evaluating the Fake News Problem at the Scale of the Information Ecosystem},
  author = {Allen, Jennifer and Howland, Baird and Mobius, Markus and Rothschild, David and Watts, Duncan J.},
  year = {2020},
  journal = {Science Advances},
  volume = {6},
  number = {14},
  issn = {2375-2548},
  doi = {10.1126/sciadv.aay3539}
}

@article{bakshy2015exposure,
  title = {Exposure to Ideologically Diverse News and Opinion on {{Facebook}}},
  author = {Bakshy, E. and Messing, S. and Adamic, L. A.},
  year = {2015},
  journal = {Science},
  volume = {348},
  number = {6239},
  pages = {1130--1132},
  issn = {0036-8075, 1095-9203},
  doi = {10.1126/science.aaa1160}
}

@inproceedings{bentley2019understanding,
  title = {Understanding Online News Behaviors},
  booktitle = {Proceedings of the 2019 {{CHI Conference}} on {{Human Factors}} in {{Computing Systems}}},
  author = {Bentley, Frank and Quehl, Katie and {Wirfs-Brock}, Jordan and Bica, Melissa},
  year = {2019},
  publisher = {{ACM}},
  address = {{Glasgow, Scotland}},
  doi = {10.1145/3290605.3300820},
  isbn = {978-1-4503-5970-2}
}

@article{bhadani2022political,
  title = {Political Audience Diversity and News Reliability in Algorithmic Ranking},
  author = {Bhadani, Saumya and Yamaya, Shun and Flammini, Alessandro and Menczer, Filippo and Ciampaglia, Giovanni Luca and Nyhan, Brendan},
  year = {2022},
  month = feb,
  journal = {Nature Human Behaviour},
  volume = {6},
  number = {4},
  pages = {495--505},
  issn = {2397-3374},
  doi = {10.1038/s41562-021-01276-5}
}

@article{brown2014political,
  title = {Political Participation of Women of Color: {{An}} Intersectional Analysis},
  author = {Brown, Nadia E.},
  year = {2014},
  journal = {Journal of Women, Politics \& Policy},
  volume = {35},
  number = {4},
  pages = {315--348},
  publisher = {{Routledge}},
  issn = {1554-477X},
  doi = {10.1080/1554477X.2014.955406}
}

@article{cardenal2019digital,
  title = {Digital Technologies and Selective Exposure: {{How}} Choice and Filter Bubbles Shape News Media Exposure},
  author = {Cardenal, Ana S. and {Aguilar-Paredes}, Carlos and Galais, Carol and {P{\'e}rez-Montoro}, Mario},
  year = {2019},
  journal = {The International Journal of Press/Politics},
  volume = {24},
  number = {4},
  pages = {465--486},
  issn = {1940-1612, 1940-1620},
  doi = {10.1177/1940161219862988}
}

@article{chen2021neutral,
  title = {Neutral Bots Probe Political Bias on Social Media},
  author = {Chen, Wen and Pacheco, Diogo and Yang, Kai-Cheng and Menczer, Filippo},
  year = {2021},
  journal = {Nature Communications},
  volume = {12},
  number = {1},
  publisher = {{Nature Publishing Group}},
  issn = {2041-1723},
  doi = {10.1038/s41467-021-25738-6},
  copyright = {2021 The Author(s)}
}

@article{cinelli2021echo,
  title = {The Echo Chamber Effect on Social Media},
  author = {Cinelli, Matteo and De Francisci Morales, Gianmarco and Galeazzi, Alessandro and Quattrociocchi, Walter and Starnini, Michele},
  year = {2021},
  journal = {Proceedings of the National Academy of Sciences},
  volume = {118},
  number = {9},
  issn = {0027-8424, 1091-6490},
  doi = {10.1073/pnas.2023301118}
}

@article{diakopoulos2015algorithmic,
  title = {Algorithmic Accountability: {{Journalistic}} Investigation of Computational Power Structures},
  author = {Diakopoulos, Nicholas},
  year = {2015},
  month = may,
  journal = {Digital Journalism},
  volume = {3},
  number = {3},
  pages = {398--415},
  issn = {2167-0811, 2167-082X},
  doi = {10.1080/21670811.2014.976411}
}

@book{ebbinghaus1913memory,
  title = {Memory: {{A}} Contribution to Experimental Psychology},
  author = {Ebbinghaus, Hermann},
  translator = {Ruger, Henry A. and Bussenius, Clara E.},
  year = {1913},
  publisher = {{Teachers College Press}},
  address = {{New York}},
  doi = {10.1037/10011-000}
}

@techreport{edelman2021edelman,
  title = {Edelman Trust Barometer 2021},
  author = {{Edelman}},
  year = {2021},
  institution = {{Edelman}}
}

@article{epstein2015search,
  title = {The Search Engine Manipulation Effect ({{SEME}}) and Its Possible Impact on the Outcomes of Elections},
  author = {Epstein, Robert and Robertson, Ronald E.},
  year = {2015},
  journal = {Proceedings of the National Academy of Sciences},
  volume = {112},
  number = {33},
  issn = {0027-8424, 1091-6490},
  doi = {10.1073/pnas.1419828112}
}

@article{epstein2017suppressing,
  title = {Suppressing the Search Engine Manipulation Effect ({{SEME}})},
  author = {Epstein, Robert and Robertson, Ronald E. and Lazer, David and Wilson, Christo},
  year = {2017},
  month = dec,
  journal = {Proceedings of the ACM on Human-Computer Interaction},
  volume = {1},
  number = {CSCW},
  pages = {1--22},
  issn = {25730142},
  doi = {10.1145/3134677}
}

@article{fischer2020auditing,
  title = {Auditing Local News Presence on {{Google News}}},
  author = {Fischer, Sean and Jaidka, Kokil and Lelkes, Yphtach},
  year = {2020},
  journal = {Nature Human Behaviour},
  volume = {4},
  number = {12},
  pages = {1236--1244},
  issn = {2397-3374},
  doi = {10.1038/s41562-020-00954-0}
}

@misc{fishkin2018new,
  title = {New Jumpshot 2018 Data: {{Where}} Searches Happen on the Web ({{Google}}, {{Amazon}}, {{Facebook}}, \& Beyond)},
  author = {Fishkin, Rand},
  year = {2018},
  month = apr,
  journal = {SparkToro},
  howpublished = {https://sparktoro.com/blog/new-jumpshot-2018-data-where-searches-happen-on-the-web-google-amazon-facebook-beyond/}
}

@article{flaxman2016filter,
  title = {Filter Bubbles, Echo Chambers, and Online News Consumption},
  author = {Flaxman, Seth and Goel, Sharad and Rao, Justin M.},
  year = {2016},
  journal = {Public Opinion Quarterly},
  volume = {80},
  number = {S1},
  pages = {298--320},
  issn = {0033-362X, 1537-5331},
  doi = {10.1093/poq/nfw006}
}

@article{fletcher2021more,
  title = {More Diverse, More Politically Varied: {{How}} Social Media, Search Engines and Aggregators Shape News Repertoires in the {{United Kingdom}}},
  author = {Fletcher, Richard and Kalogeropoulos, Antonis and Nielsen, Rasmus Kleis},
  year = {2021},
  month = jul,
  journal = {New Media \& Society},
  issn = {1461-4448, 1461-7315},
  doi = {10.1177/14614448211027393}
}

@article{garimella2021political,
  title = {Political Polarization in Online News Consumption},
  author = {Garimella, Kiran and Smith, Tim and Weiss, Rebecca and West, Robert},
  year = {2021},
  month = may,
  journal = {Proceedings of the International AAAI Conference on Web and Social Media},
  volume = {15},
  pages = {152--162},
  issn = {2334-0770},
  copyright = {Copyright (c) 2021 Association for the Advancement of Artificial Intelligence}
}

@techreport{golebiewski2019data,
  title = {Data Voids: {{Where}} Missing Data Can Easily Be Exploited},
  author = {Golebiewski, Michael and {boyd}, danah},
  year = {2019},
  institution = {{Data \& Society}}
}

@article{grinberg2019fake,
  title = {Fake News on {{Twitter}} during the 2016 {{U}}.{{S}}. Presidential Election},
  author = {Grinberg, Nir and Joseph, Kenneth and Friedland, Lisa and {Swire-Thompson}, Briony and Lazer, David},
  year = {2019},
  journal = {Science},
  volume = {363},
  number = {6425},
  pages = {374--378},
  issn = {0036-8075, 1095-9203},
  doi = {10.1126/science.aau2706}
}

@article{guess2019less,
  title = {Less than You Think: {{Prevalence}} and Predictors of Fake News Dissemination on {{Facebook}}},
  author = {Guess, Andrew M. and Nagler, Jonathan and Tucker, Joshua},
  year = {2019},
  journal = {Science Advances},
  volume = {5},
  number = {1},
  issn = {2375-2548},
  doi = {10.1126/sciadv.aau4586}
}

@article{guess2020exposure,
  title = {Exposure to Untrustworthy Websites in the 2016 {{US}} Election},
  author = {Guess, Andrew M. and Nyhan, Brendan and Reifler, Jason},
  year = {2020},
  journal = {Nature Human Behaviour},
  volume = {4},
  number = {5},
  pages = {472--480},
  issn = {2397-3374},
  doi = {10.1038/s41562-020-0833-x}
}

@article{guess2021almost,
  title = {({{Almost}}) Everything in Moderation: {{New}} Evidence on {{Americans}}' Online Media Diets},
  author = {Guess, Andrew M.},
  year = {2021},
  journal = {American Journal of Political Science},
  volume = {65},
  number = {4},
  pages = {1007--1022},
  issn = {0092-5853, 1540-5907},
  doi = {10.1111/ajps.12589}
}

@article{guess2021consequences,
  title = {The Consequences of Online Partisan Media},
  author = {Guess, Andrew M. and Barber{\'a}, Pablo and Munzert, Simon and Yang, JungHwan},
  year = {2021},
  month = apr,
  journal = {Proceedings of the National Academy of Sciences},
  volume = {118},
  number = {14},
  issn = {0027-8424, 1091-6490},
  doi = {10.1073/pnas.2013464118}
}

@inproceedings{hannak2013measuring,
  title = {Measuring Personalization of Web Search},
  booktitle = {Proceedings of the 22nd {{International Conference}} on {{World Wide Web}}},
  author = {Hannak, Aniko and Sapiezynski, Piotr and Molavi Kakhki, Arash and Krishnamurthy, Balachander and Lazer, David and Mislove, Alan and Wilson, Christo},
  year = {2013},
  pages = {527--538},
  doi = {10.1145/2488388.2488435},
  isbn = {978-1-4503-2035-1}
}

@misc{hobbs2019text,
  type = {{{SSRN Scholarly Paper}}},
  title = {Text Scaling for Open-Ended Survey Responses and Social Media Posts},
  author = {Hobbs, William R.},
  year = {2019},
  month = aug,
  number = {3044864},
  address = {{Rochester, NY}},
  doi = {10.2139/ssrn.3044864}
}

@article{hosseinmardi2021examining,
  title = {Examining the Consumption of Radical Content on {{YouTube}}},
  author = {Hosseinmardi, Homa and Ghasemian, Amir and Clauset, Aaron and Mobius, Markus and Rothschild, David M. and Watts, Duncan J.},
  year = {2021},
  journal = {Proceedings of the National Academy of Sciences},
  volume = {118},
  number = {32},
  issn = {0027-8424, 1091-6490},
  doi = {10.1073/pnas.2101967118}
}

@article{huszar2022algorithmic,
  title = {Algorithmic Amplification of Politics on {{Twitter}}},
  author = {Husz{\'a}r, Ferenc and Ktena, Sofia Ira and O'Brien, Conor and Belli, Luca and Schlaikjer, Andrew and Hardt, Moritz},
  year = {2022},
  month = jan,
  journal = {Proceedings of the National Academy of Sciences},
  volume = {119},
  number = {1},
  publisher = {{National Academy of Sciences}},
  issn = {0027-8424, 1091-6490},
  doi = {10.1073/pnas.2025334119},
  chapter = {Social Sciences},
  copyright = {Copyright \textcopyright{} 2021 the Author(s). Published by PNAS.. https://creativecommons.org/licenses/by-nc-nd/4.0/This open access article is distributed under Creative Commons Attribution-NonCommercial-NoDerivatives License 4.0 (CC BY-NC-ND).},
  pmid = {34934011}
}

@article{hyde2019future,
  title = {The Future of Sex and Gender in Psychology: {{Five}} Challenges to the Gender Binary},
  author = {Hyde, Janet Shibley and Bigler, Rebecca S. and Joel, Daphna and Tate, Charlotte Chucky and {van Anders}, Sari M.},
  year = {2019},
  journal = {American Psychologist},
  volume = {74},
  number = {2},
  pages = {171--193},
  publisher = {{American Psychological Association}},
  address = {{US}},
  issn = {1935-990X},
  doi = {10.1037/amp0000307}
}

@article{introna2000shaping,
  title = {Shaping the Web: {{Why}} the Politics of Search Engines Matters},
  author = {Introna, Lucas D. and Nissenbaum, Helen},
  year = {2000},
  journal = {The Information Society},
  volume = {16},
  number = {3},
  pages = {169--185},
  issn = {0197-2243, 1087-6537},
  doi = {10.1080/01972240050133634}
}

@article{iyengar2008selective,
  title = {Selective Exposure to Campaign Communication: {{The}} Role of Anticipated Agreement and Issue Public Membership},
  author = {Iyengar, Shanto and Hahn, Kyu S. and Krosnick, Jon A. and Walker, John},
  year = {2008},
  journal = {The Journal of Politics},
  volume = {70},
  number = {1},
  pages = {186--200},
  issn = {0022-3816, 1468-2508},
  doi = {10.1017/S0022381607080139}
}

@article{iyengar2009red,
  title = {Red Media, Blue Media: {{Evidence}} of Ideological Selectivity in Media Use},
  author = {Iyengar, Shanto and Hahn, Kyu S},
  year = {2009},
  journal = {Journal of Communication},
  volume = {59},
  number = {1},
  pages = {19--39},
  issn = {00219916, 14602466},
  doi = {10.1111/j.1460-2466.2008.01402.x}
}

@article{johnson2020online,
  title = {The Online Competition between Pro- and Anti-Vaccination Views},
  author = {Johnson, Neil F. and Vel{\'a}squez, Nicolas and Restrepo, Nicholas Johnson and Leahy, Rhys and Gabriel, Nicholas and El Oud, Sara and Zheng, Minzhang and Manrique, Pedro and Wuchty, Stefan and Lupu, Yonatan},
  year = {2020},
  journal = {Nature},
  volume = {582},
  number = {7811},
  pages = {230--233},
  publisher = {{Nature Publishing Group}},
  issn = {1476-4687},
  doi = {10.1038/s41586-020-2281-1},
  copyright = {2020 The Author(s), under exclusive licence to Springer Nature Limited}
}

@inproceedings{kawakami2020media,
  title = {The Media Coverage of the 2020 {{US}} Presidential Election Candidates through the Lens of {{Google}}'s Top Stories},
  booktitle = {Proceedings of the {{International AAAI Conference}} on {{Web}} and {{Social Media}}},
  author = {Kawakami, Anna and Umarova, Khonzodakhon and Mustafaraj, Eni},
  year = {2020},
  month = may,
  volume = {14},
  pages = {868--877},
  copyright = {Copyright (c) 2020 Association for the Advancement of Artificial Intelligence}
}

@book{klar2016independent,
  title = {Independent Politics: {{How American}} Disdain for Parties Leads to Political Inaction},
  author = {Klar, Samara and Krupnikov, Yanna},
  year = {2016},
  publisher = {{Cambridge University Press}},
  address = {{New York, NY}},
  isbn = {978-1-107-13446-1},
  lccn = {JK2271 .K53 2016}
}

@article{kulshrestha2019search,
  title = {Search Bias Quantification: {{Investigating}} Political Bias in Social Media and Web Search},
  author = {Kulshrestha, Juhi and Eslami, Motahhare and Messias, Johnnatan and Zafar, Muhammad Bilal and Ghosh, Saptarshi and Gummadi, Krishna P. and Karahalios, Karrie},
  year = {2019},
  journal = {Information Retrieval Journal},
  volume = {22},
  number = {1},
  pages = {188--227},
  issn = {1386-4564, 1573-7659},
  doi = {10.1007/s10791-018-9341-2}
}

@article{lawrence1999accessibility,
  title = {Accessibility of Information on the Web},
  author = {Lawrence, Steve and Giles, C. Lee},
  year = {1999},
  journal = {Nature},
  volume = {400},
  number = {6740},
  pages = {107--107},
  issn = {0028-0836, 1476-4687},
  doi = {10.1038/21987}
}

@article{lazer2018science,
  title = {The Science of Fake News},
  author = {Lazer, David M. J. and Baum, Matthew A. and Benkler, Yochai and Berinsky, Adam J. and Greenhill, Kelly M. and Menczer, Filippo and Metzger, Miriam J. and Nyhan, Brendan and Pennycook, Gordon and Rothschild, David and Schudson, Michael and Sloman, Steven A. and Sunstein, Cass R. and Thorson, Emily A. and Watts, Duncan J. and Zittrain, Jonathan L.},
  year = {2018},
  journal = {Science},
  volume = {359},
  number = {6380},
  pages = {1094--1096},
  issn = {0036-8075, 1095-9203},
  doi = {10.1126/science.aao2998}
}

@article{lewandowsky2017misinformation,
  title = {Beyond Misinformation: {{Understanding}} and Coping with the ``Post-Truth'' Era},
  author = {Lewandowsky, Stephan and Ecker, Ullrich K. H. and Cook, John},
  year = {2017},
  month = dec,
  journal = {Journal of Applied Research in Memory and Cognition},
  volume = {6},
  number = {4},
  pages = {353--369},
  issn = {2211-369X, 2211-3681},
  doi = {10.1016/j.jarmac.2017.07.008}
}

@article{metaxa2019search,
  title = {Search Media and Elections: {{A}} Longitudinal Investigation of Political Search Results},
  author = {Metaxa, Dana{\"e} and Park, Joon Sung and Landay, James A. and Hancock, Jeff},
  year = {2019},
  journal = {Proceedings of the ACM on Human-Computer Interaction},
  volume = {3},
  number = {CSCW},
  pages = {1--17},
  issn = {25730142},
  doi = {10.1145/3359231}
}

@inproceedings{metaxas2005weba,
  title = {Web Spam, Propaganda and Trust},
  booktitle = {Proceedings of the 2005 {{World Wide Web Conference}}},
  author = {Metaxas, Panagiotis T and DeStefano, Joseph},
  year = {2005}
}

@techreport{mitchell2017how,
  title = {How {{Americans}} Encounter, Recall and Act upon Digital News},
  author = {Mitchell, Amy and Gottfried, Jeffrey and Shearer, Elisa and Lu, Kristine},
  year = {2017},
  institution = {{Pew Research Center}}
}

@article{muise2022quantifying,
  title = {Quantifying Partisan News Diets in {{Web}} and {{TV}} Audiences},
  author = {Muise, Daniel and Hosseinmardi, Homa and Howland, Baird and Mobius, Markus and Rothschild, David and Watts, Duncan J.},
  year = {2022},
  month = jul,
  journal = {Science Advances},
  volume = {8},
  number = {28},
  publisher = {{American Association for the Advancement of Science}},
  doi = {10.1126/sciadv.abn0083}
}

@inproceedings{mustafaraj2020case,
  title = {The Case for Voter-Centered Audits of Search Engines during Political Elections},
  booktitle = {Proceedings of the 2020 {{Conference}} on {{Fairness}}, {{Accountability}}, and {{Transparency}}},
  author = {Mustafaraj, Eni and Lurie, Emma and Devine, Claire},
  year = {2020},
  pages = {559--569}
}

@article{nanz2022democratic,
  title = {Democratic Consequences of Incidental Exposure to Political Information: {{A}} Meta-Analysis},
  author = {Nanz, Andreas and Matthes, J{\"o}rg},
  year = {2022},
  month = mar,
  journal = {Journal of Communication},
  volume = {72},
  number = {3},
  pages = {345--373},
  issn = {0021-9916},
  doi = {10.1093/joc/jqac008}
}

@techreport{newman2019digital,
  title = {Digital News Report 2019},
  author = {Newman, Nic and Fletcher, Richard and Kalogeropoulos, Antonis and Neilsen, Rasmus Kleis},
  year = {2019},
  institution = {{Reuters Institute for the Study of Journalism}}
}

@misc{newsguard2021rating,
  title = {Rating Process and Criteria},
  author = {NewsGuard},
  year = {2021},
  journal = {Rating Process and Criteria},
  howpublished = {https://www.newsguardtech.com/ratings/rating-process-criteria/}
}

@book{noble2018algorithms,
  title = {Algorithms of Oppression: {{How}} Search Engines Reinforce Racism},
  author = {Noble, Safiya Umoja},
  year = {2018},
  publisher = {{New York University Press}},
  address = {{New York}},
  isbn = {978-1-4798-4994-9},
  lccn = {ZA4230 .N63 2018}
}

@article{pan2007google,
  title = {In {{Google}} We Trust: {{Users}}' Decisions on Rank, Position, and Relevance},
  author = {Pan, Bing and Hembrooke, Helene and Joachims, Thorsten and Lorigo, Lori and Gay, Geri and Granka, Laura},
  year = {2007},
  journal = {Journal of Computer-Mediated Communication},
  volume = {12},
  number = {3},
  pages = {801--823},
  issn = {10836101, 10836101},
  doi = {10.1111/j.1083-6101.2007.00351.x}
}

@book{pariser2011filter,
  title = {The Filter Bubble: {{What}} the Internet Is Hiding from You},
  author = {Pariser, Eli},
  year = {2011},
  publisher = {{Penguin}},
  address = {{New York, NY}},
  isbn = {978-0-14-312123-7}
}

@article{pennycook2021shifting,
  title = {Shifting Attention to Accuracy Can Reduce Misinformation Online},
  author = {Pennycook, Gordon and Epstein, Ziv and Mosleh, Mohsen and Arechar, Antonio A. and Eckles, Dean and Rand, David G.},
  year = {2021},
  month = apr,
  journal = {Nature},
  volume = {592},
  number = {7855},
  pages = {590--595},
  publisher = {{Nature Publishing Group}},
  issn = {1476-4687},
  doi = {10.1038/s41586-021-03344-2},
  copyright = {2021 The Author(s), under exclusive licence to Springer Nature Limited}
}

@article{peterson2021partisan,
  title = {Partisan Selective Exposure in Online News Consumption: {{Evidence}} from the 2016 Presidential Campaign},
  author = {Peterson, Erik and Goel, Sharad and Iyengar, Shanto},
  year = {2021},
  month = apr,
  journal = {Political Science Research and Methods},
  volume = {9},
  number = {2},
  pages = {242--258},
  issn = {2049-8470, 2049-8489},
  doi = {10.1017/psrm.2019.55}
}

@article{rahwan2019machine,
  title = {Machine Behaviour},
  author = {Rahwan, Iyad and Cebrian, Manuel and Obradovich, Nick and Bongard, Josh and Bonnefon, Jean-Fran{\c c}ois and Breazeal, Cynthia and Crandall, Jacob W. and Christakis, Nicholas A. and Couzin, Iain D. and Jackson, Matthew O. and Jennings, Nicholas R. and Kamar, Ece and Kloumann, Isabel M. and Larochelle, Hugo and Lazer, David and McElreath, Richard and Mislove, Alan and Parkes, David C. and Pentland, Alex `Sandy' and Roberts, Margaret E. and Shariff, Azim and Tenenbaum, Joshua B. and Wellman, Michael},
  year = {2019},
  journal = {Nature},
  volume = {568},
  number = {7753},
  pages = {477--486},
  issn = {0028-0836, 1476-4687},
  doi = {10.1038/s41586-019-1138-y}
}

@article{reeves2021screenomics,
  title = {Screenomics: {{A}} Framework to Capture and Analyze Personal Life Experiences and the Ways That Technology Shapes Them},
  author = {Reeves, Byron and Ram, Nilam and Robinson, Thomas N. and Cummings, James J. and Giles, C. Lee and Pan, Jennifer and Chiatti, Agnese and Cho, Mj and Roehrick, Katie and Yang, Xiao and Gagneja, Anupriya and Brinberg, Miriam and Muise, Daniel and Lu, Yingdan and Luo, Mufan and Fitzgerald, Andrew and Yeykelis, Leo},
  year = {2021},
  month = mar,
  journal = {Human\textendash Computer Interaction},
  volume = {36},
  number = {2},
  pages = {150--201},
  issn = {0737-0024, 1532-7051},
  doi = {10.1080/07370024.2019.1578652}
}

@article{robertson2018auditinga,
  title = {Auditing Partisan Audience Bias within {{Google Search}}},
  author = {Robertson, Ronald E. and Jiang, Shan and Joseph, Kenneth and Friedland, Lisa and Lazer, David and Wilson, Christo},
  year = {2018},
  journal = {Proceedings of the ACM on Human-Computer Interaction},
  volume = {2},
  number = {CSCW},
  pages = {1--22},
  issn = {25730142},
  doi = {10.1145/3274417}
}

@misc{statcounter2020desktop,
  title = {Desktop Search Engine Market Share {{United States}} of {{America}}},
  author = {StatCounter},
  year = {2020},
  journal = {StatCounter Global Stats},
  howpublished = {https://gs.statcounter.com/search-engine-market-share/desktop/united-states-of-america/2020}
}

@article{stier2021post,
  title = {Post Post-Broadcast Democracy? {{News}} Exposure in the Age of Online Intermediaries},
  author = {Stier, Sebastian and Mangold, Frank and Scharkow, Michael and Breuer, Johannes},
  year = {2021},
  journal = {American Political Science Review},
  volume = {116},
  number = {2},
  pages = {768--774},
  publisher = {{Cambridge University Press}},
  issn = {0003-0554, 1537-5943},
  doi = {10.1017/S0003055421001222}
}

@book{sunstein2001republic,
  title = {Republic{{.com}}},
  author = {Sunstein, Cass R.},
  year = {2001},
  publisher = {{Princeton University Press}},
  googlebooks = {O7AG9TxDJdgC},
  isbn = {978-0-691-09589-9}
}

@article{taber2006motivated,
  title = {Motivated Skepticism in the Evaluation of Political Beliefs},
  author = {Taber, Charles S. and Lodge, Milton},
  year = {2006},
  journal = {American Journal of Political Science},
  volume = {50},
  number = {3},
  pages = {755--769},
  issn = {0092-5853, 1540-5907},
  doi = {10.1111/j.1540-5907.2006.00214.x}
}

@inproceedings{trielli2019search,
  title = {Search as News Curator: {{The}} Role of {{Google}} in Shaping Attention to News Information},
  booktitle = {Proceedings of the 2019 {{CHI Conference}} on {{Human Factors}} in {{Computing Systems}}},
  author = {Trielli, Daniel and Diakopoulos, Nicholas},
  year = {2019},
  publisher = {{ACM Press}},
  address = {{Glasgow, Scotland Uk}},
  doi = {10.1145/3290605.3300683},
  isbn = {978-1-4503-5970-2}
}

@article{trielli2020partisan,
  title = {Partisan Search Behavior and {{Google}} Results in the 2018 {{U}}.{{S}}. Midterm Elections},
  author = {Trielli, Daniel and Diakopoulos, Nicholas},
  year = {2020},
  journal = {Information, Communication \& Society},
  volume = {25},
  pages = {145--161},
  issn = {1369-118X, 1468-4462},
  doi = {10.1080/1369118X.2020.1764605}
}

@book{vaidhyanathan2011googlization,
  title = {The {{Googlization}} of Everything (and Why We Should Worry)},
  author = {Vaidhyanathan, Siva},
  year = {2011},
  publisher = {{University of California press}},
  address = {{Berkeley, CA}},
  isbn = {978-0-520-25882-2},
  lccn = {025.04}
}

@article{vanhoof2022searching,
  title = {Searching Differently? {{How}} Political Attitudes Impact Search Queries about Political Issues},
  author = {{van Hoof}, Marieke and Meppelink, Corine S and Moeller, Judith and Trilling, Damian},
  year = {2022},
  month = jul,
  journal = {New Media \& Society},
  publisher = {{SAGE Publications}},
  issn = {1461-4448},
  doi = {10.1177/14614448221104405}
}

@article{wagner2021measuring,
  title = {Measuring Algorithmically Infused Societies},
  author = {Wagner, Claudia and Strohmaier, Markus and Olteanu, Alexandra and K{\i}c{\i}man, Emre and Contractor, Noshir and {Eliassi-Rad}, Tina},
  year = {2021},
  journal = {Nature},
  volume = {595},
  number = {7866},
  pages = {197--204},
  issn = {0028-0836, 1476-4687},
  doi = {10.1038/s41586-021-03666-1}
}

@techreport{walker2019americans,
  title = {Americans Favor Mobile Devices over Desktops and Laptops for Getting News},
  author = {Walker, Mason},
  year = {2019},
  institution = {{Pew Research Center}}
}

@article{watts2021measuring,
  title = {Measuring the News and Its Impact on Democracy},
  author = {Watts, Duncan J. and Rothschild, David M. and Mobius, Markus},
  year = {2021},
  journal = {Proceedings of the National Academy of Sciences},
  volume = {118},
  number = {15},
  publisher = {{National Academy of Sciences}},
  issn = {0027-8424, 1091-6490},
  doi = {10.1073/pnas.1912443118},
  chapter = {Colloquium Paper},
  copyright = {\textcopyright{} 2021 . https://www.pnas.org/site/aboutpnas/licenses.xhtmlPublished under the PNAS license.},
  pmid = {33837145}
}

@article{wojcieszak2021avenues,
  title = {Avenues to News and Diverse News Exposure Online: {{Comparing}} Direct Navigation, Social Media, News Aggregators, Search Queries, and Article Hyperlinks},
  author = {Wojcieszak, Magdalena and {Menchen-Trevino}, Ericka and Goncalves, Joao F. F. and Weeks, Brian},
  year = {2021},
  journal = {The International Journal of Press/Politics},
  issn = {1940-1612, 1940-1620},
  doi = {10.1177/19401612211009160}
}

@misc{yin2018local,
  title = {Local News Dataset},
  author = {Yin, Leon},
  year = {2018},
  doi = {10.5281/ZENODO.1345145},
  copyright = {Open Access},
  howpublished = {Zenodo}
}

@article{zuckerman2021why,
  title = {Why Study Media Ecosystems?},
  author = {Zuckerman, Ethan},
  year = {2021},
  journal = {Information, Communication \& Society},
  volume = {24},
  number = {10},
  pages = {1--19},
  issn = {1369-118X, 1468-4462},
  doi = {10.1080/1369118X.2021.1942513}
}
